\documentclass[fleqn,usenatbib]{mnras}

\usepackage{newtxtext,newtxmath}

\usepackage[T1]{fontenc}

\usepackage{graphicx}	
\usepackage{amsmath}	
\usepackage{color,appendix,soul}

\newcommand{\kms}{${\rm\,km\,s^{-1}}$} 
\newcommand{\gppr}{\stackrel{>}{\scriptstyle \sim}}
\newcommand{\gappr}{\raisebox{-0.4ex}{$\gppr$}}
\newcommand{\lppr}{\stackrel{<}{\scriptstyle \sim}}
\newcommand{\lappr}{\raisebox{-0.4ex}{$\lppr$}}

\title[The {\it Gaia} DR2 halo white dwarf population]{The {\it Gaia} DR2 halo white dwarf population: the luminosity function, mass distribution and its star formation history}

\author[S. Torres et al.]{
Santiago Torres$^{1,2}$,\thanks{E-mail: santiago.torres@upc.edu}
Alberto Rebassa-Mansergas$^{1,2}$,
Mar\'ia E. Camisassa$^{1,3,4}$ and \newauthor
Roberto Raddi$^{1}$
\\
$^{1}$ Departament de F\'\i sica, 
           Universitat Polit\`ecnica de Catalunya, 
           c/Esteve Terrades 5, 
           08860 Castelldefels, 
           Spain\\
$^{2}$ Institute for Space Studies of Catalonia, 
           c/Gran Capita 2--4, 
           Edif. Nexus 104, 
           08034 Barcelona, 
           Spain\\
$^{3}$Facultad de Ciencias Astron\'omicas y Geof\'{\i}sicas, 
           Universidad Nacional de La Plata, 
           Paseo del Bosque s/n, 1900 
           La Plata, 
           Argentina\\
$^{4}$ Instituto de Astrof\'isica de La Plata, UNLP-CONICET,
           Paseo del Bosque s/n, 
           1900 La Plata, 
           Argentina
}

\date{Accepted XXX. Received YYY; in original form ZZZ}

\pubyear{2015}

\begin{document}
\label{firstpage}
\pagerange{\pageref{firstpage}--\pageref{lastpage}}
\maketitle

\begin{abstract}
We analyze the volume-limited nearly complete 100\,pc sample of 95 halo white dwarf candidates identified by the second data release of {\it Gaia}. Based on a detailed population synthesis model, we apply a method that relies on {\it Gaia} astrometry and photometry to accurately derive the individual white dwarf parameters (mass, radius, effective temperature, bolometric luminosity and age). This method is tested with 25 white dwarfs of our sample for which we took optical spectra and performed spectroscopic analysis. We build and analyse the halo white dwarf luminosity function, for which we find for the first time possible evidences of the  cut-off at its faintest end, leading to an age estimate of $\simeq12\pm0.5\,$Gyr. The mass distribution of the sample peaks at $0.589\,M_{\odot}$, with $71\%$ of the white dwarf masses below $0.6\,M_{\odot}$ and just two massive white dwarfs of more than $0.8\,M_{\odot}$. From the age distribution we find three white dwarfs with total ages above 12\,Gyr, of which J1312-4728 is the oldest white dwarf known with an age of $12.41\pm0.22\,$Gyr. We prove that the star formation history is mainly characterised by a burst of star formation that occurred from 10 to 12\,Gyr in the past, but extended up to 8 Gyr. We also find that the peak of the star formation history is centered at around 11\,Gyr, which is compatible with the current age of the Gaia-Enceladus encounter. Finally, $13\%$ of our halo sample is contaminated by high-speed young objects (total age<7\,Gyr). The origin of these white dwarfs is unclear but their age distribution may be compatible with the encounter with the Sagittarius galaxy.
\end{abstract}




White dwarfs are long-lived objects whose evolutionary characteristics are reasonably well understood \citep[e.g.][and references therein]{Althaus2010}. They represent the vast majority of low- and intermediate-mass stars remnants. Thus, their ensemble properties carry valuable information about the past history and evolution of the different components of the Galaxy. In particular, white dwarfs are reliable cosmochronometers and, consequently, they have been used for studying several Galactic populations. As illustrative examples of this, we can mention the analysis of the Galactic thin and thick disk \cite[e.g.][]{GBerro1999,Torres2002,Rowell2011,Kilic2017}, the halo \cite[e.g.][]{Mochkovitch1990,Isern1998,GBerro2004,Cojocaru2015} the bulge \citep[e.g][]{Calamida2014,Torres2018} or precise studies of Galactic open and globular
clusters \cite[e.g.][]{GBerro2010,Jeffery2011,Hansen2013,Torres2015}.

Regarding the population of white dwarfs in the Galactic stellar halo, it has been the focus of attention in this field since the first observational and theoretical studies \citep{Mochkovitch1990,Liebert1989,Isern1998}. Shortly after, and since the MACHO collaboration experiment for the microlensing detection \citep[e.g.][]{Alcock2000}, halo white dwarfs have been suggested as natural candidates to contribute to the dark matter content of the Galaxy \citep[e.g.][]{Oppenheimer2001}. An intense debate arose on this issue,  although later studies demonstrated that the white dwarf contribution to dark matter was rather limited \citep[e.g.][]{Torres2002,Flynn2003,GBerro2004,Kilic2004,Bergeron2005}. In any case, the search for white dwarfs in the Galactic halo has proven to be a difficult endeavour. In this sense, the intrinsic faintness and the low space density of halo white dwarfs, along with their high surface gravity, which erases any trace on metal content from their atmospheres and disables accurate radial velocity measurements due to the broadening of the Balmer lines, are some of the main factors that have hindered the discovery of suitable candidates. Consequently, during  decades, the identification of halo members has relied  on the search of cool objects in high proper motion surveys \citep[e.g.][]{Monet2000,Nelson2002,Hall2008,Catalan2012,Munn2016}. It was not until the ESO SNIa Progenitor
surveY (SPY) project \citep[see][and references therein]{Napiwotzki2020} that radial velocities were measured for the first time with enough precision to recover the 3D kinematics of white dwarfs. That permitted to identify high eccentric retrograde orbits as strongly indicators of their belonging to the halo population \citep{Pauli2006}. Although the number of halo members was sparse, an estimate of the age of the inner halo has been possible \citep[$\sim11.4\,$Gyr;][]{Kalirai2012}.

A key ingredient of these studies is the white dwarf  luminosity function. Defined as the number of white dwarfs per bolometric magnitude unit and cubic parsec, the white dwarf luminosity function is a valuable tool to derive the age, history and evolution of the components of our Galaxy. Moreover, the luminosity function holds all the information about the cooling process of the white dwarfs, being then  an excellent tool for testing the physics of evolutionary models (see \citealt{GBerroOswalt2016}, for a comprehensive review about these issues). However, the effectiveness of the white dwarf luminosity function requires the existence of a complete,
well-defined volume sample. In this sense, the lack of accurate distances previous to the {\it Gaia} era, joint to the selection biases inherited from magnitude-limited samples, have hampered the achievement of more conclusive results in such an elusive sample as the halo white dwarf population.

\begin{figure*}
      {\includegraphics[width=1.0\columnwidth, clip=true,trim=5 15 20 30]{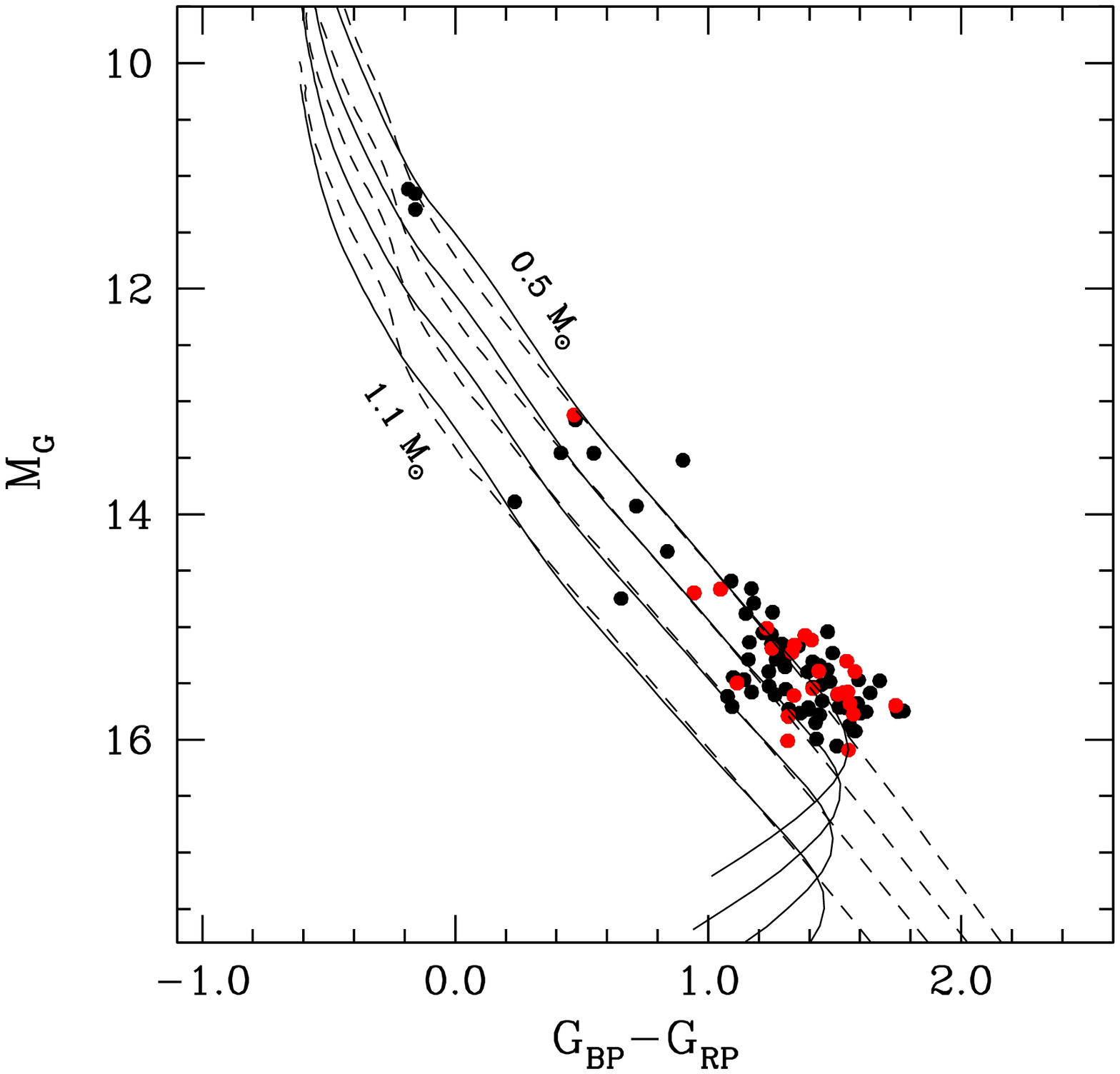}}
        {\includegraphics[width=1.0\columnwidth,clip=true,trim=5 15 20 30]{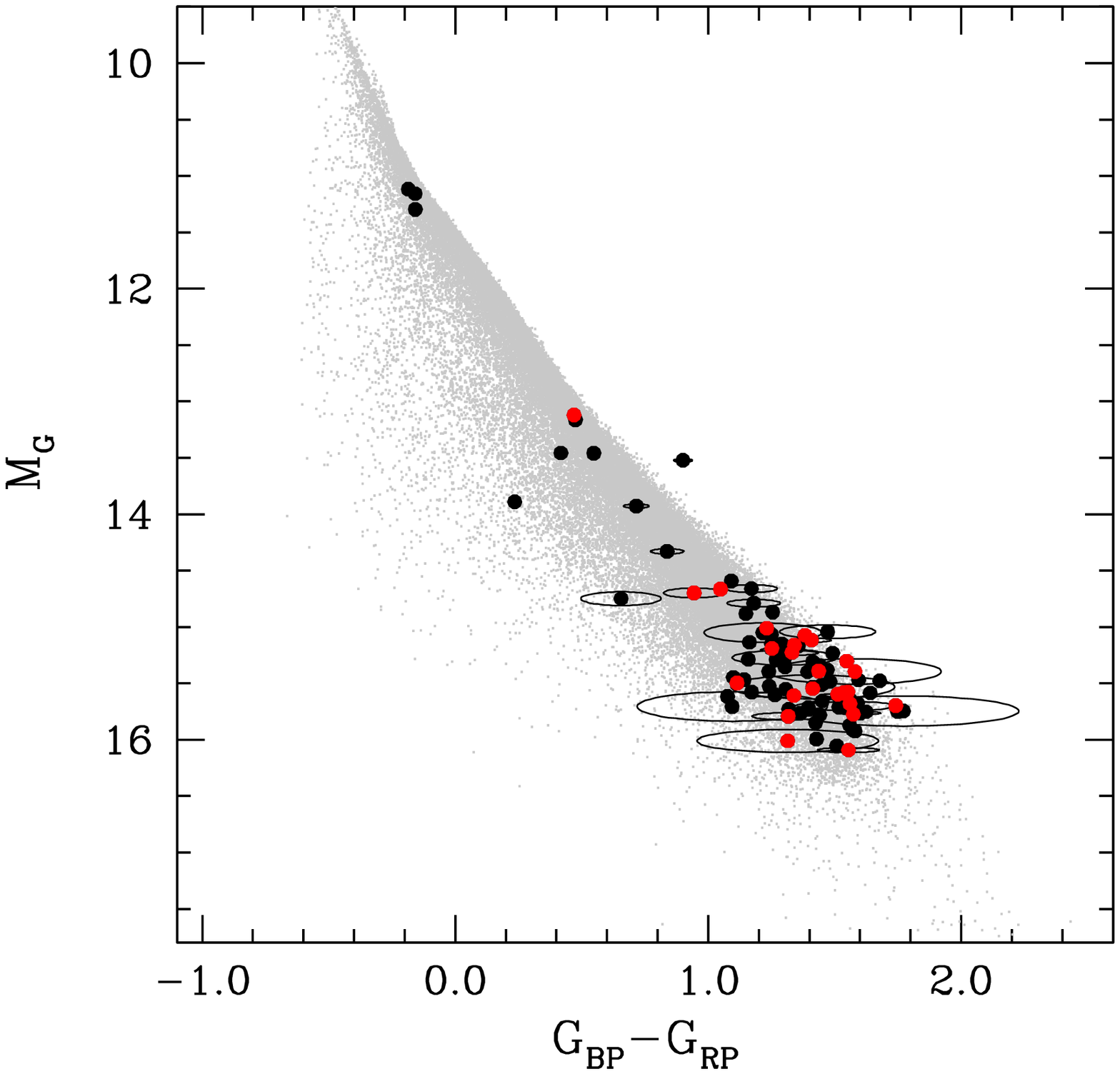}}  
  \caption{{\it Gaia} HR-diagram for our halo white dwarf sample within 100\,pc. Marked as red circles are those white dwarfs for which we obtained optical spectra. In the left panel we plot the theoretical cooling tracks for DA (solid lines) and DB (dashed lines) white dwarfs for different masses ranging from 0.5 to $1.1\,M_{\odot}$ in steps of $0.2\,M_{\odot}$. In the right panel a typical simulation (gray dots) taking into account photometric and astrometric errors is shown.  For illustrative purposes and for the sake of clarity  we only plot some $1\sigma$ ellipses. See text for details. } 
  \label{f:HR-tracks}
\end{figure*}

The first observational halo white dwarf luminosity function \citep{Liebert1989} contained only five objects, and no more than 20 were proposed as halo candidates through a neural network artificial intelligent classification algorithm \citep{Torres1998}. Based on the SuperCOSMOS Sky Survey observations towards the southern  Galactic Cap, \cite{Oppenheimer2001} identified 38 high-tangential velocity white dwarfs, from which they deduced that white dwarfs may represent 2\% of the local dark matter density. An intense debate began in this regard, mainly based on the difficulty of unequivocally identifying halo members. Currently, several halo white dwarf luminosity functions have been proposed \citep[e.g.][]{Rowell2011,Munn2016,Lam2018}. These samples, based on reduced-proper motion surveys, are magnitude limited. Thus, different statistical approaches, starting from  the well-known maximum volume estimator method, have been applied in order to correct the observational biases of the intrinsic  incompleteness of magnitude limited samples. Parallel to the search for statistical significant halo white dwarf samples, a major effort has been devoted to identify individual halo white dwarf members. The list includes from hot and warm objects up to cool and ultracool white dwarfs \citep[e.g.][]{Hambly1997,Ibata2000,Pauli2006,Kalirai2012,Kawka2012,Gianninas2015,Si2017}. 

However, as stated before, achieving a statistical complete sample of halo white dwarfs has become an arduous task, at least during the pre-{\it Gaia} era. A first estimate of the number of halo white dwarfs showed that up to magnitude $G<20$, around $85$ objects within 100\,pc and $\sim500$ objects within 400\,pc are expected to be accessible by {\it Gaia} \citep{Torres2005}. Since {\it Gaia} Data Release 2 was published only a few studies of the Galactic halo white dwarf population have been published. In \cite{Kilic2018}, a sample of 142 objects inconsistent with disk kinematics has been analyzed. In particular, an age estimate has been provided for many of the members of the sample, obtaining, from the coolest white dwarfs, an age estimate of the inner halo of 10.9\,Gyr. However, the sample, which contains objects as far as 540\,pc, is far from being complete. On the other hand, \cite{Torres2019} focused on the search of halo white dwarfs in the near-complete sample within 100\,pc from the Sun. With the help of artificial intelligence algorithms, \cite{Torres2019} categorized the membership of white dwarfs in the different Galactic components, identifying the largest volume-complete halo white dwarf sample, consisting in 95 stars.

A renewed interest on the halo population has appeared since the astrometric {\it Gaia} mission has provided accurate parallax and proper motion measurements for approximately 1.4 billion stars of our Galaxy \citep{Gaiacol2018}. In the new Gaia-Enceladus paradigm, a major impact event took place in our Galaxy around 10.5-11.5\,Gyr in the past, being at the origin of the formation of the thick disk and inner stellar halo \citep{Helmi2018,Gallart2019}. The precise age, intensity and effects of this past event on the kinematics and properties of the stars in the solar neighborhood are not yet  well understood. 

In this paper, we  analyze the halo white dwarf sample identified in \cite{Torres2019}. In particular, we obtain the halo white dwarf luminosity function, the mass distribution of the sample and an estimate of its star formation history. To achieve this goal, we complement our analysis  with spectroscopic observation of 27 stars of the sample. The stellar parameters of each white dwarf (i.e.  luminosity, age and mass) are obtained from a robust method which makes use of {\it Gaia} photometry and astrometry together with a detailed population synthesis code, based on Monte Carlo techniques and which incorporates the most up-to-date evolutionary sequences of white dwarfs. Finally, it is  important to emphasise  that we use the term halo to describe that sample whose characteristics are different from the average thin or thick disk white dwarf sample. If these objects really belong to an ancient halo, inner spheroid or are the remnants of a major merger event, it is something that will be analyzed along the paper.

The paper is organized as follows. In the first section we describe our halo white dwarf sample. That includes how we identified the halo candidates, the fundamental parameters that were derived from those spectroscopically observed and the main characteristics of our population synthesis modeling. In the second section we present our strategy and testing for deriving the fundamental parameters extended to the whole halo white dwarf sample. Then, the halo white dwarf luminosity function, mass distribution and star formation history are presented. Finally, in Section \ref{conclusions} we analyze the results achieved and summarize them in the concluding remarks.

\section{The halo white dwarf sample}

\subsection{Identification of halo white dwarfs}
\label{ss:ide} 

{\it Gaia} Data Release 2 has provided a wealth of unprecedented information concerning the Galactic white dwarf population \citep{Jimenez2018,Fusillo2019}. The high astrometric accuracy and the photometry provided by {\it Gaia} has allowed us to build a clean
color-magnitude diagram to select a large sample of white dwarf candidates. With the aid of a detailed population synthesis simulator, we analysed the main properties of the white dwarf  population available from {\it Gaia} \citep{Jimenez2018}. In this study, we showed that the largest and most complete sample of white dwarfs available by {\it Gaia} is up to 100 pc from the Sun. For larger distances, the completeness of the sample decays dramatically and biases and magnitude-selection effects begin to grow in importance.

With the aid of advanced intelligent algorithms, we studied the main properties of the Gaia 100 pc white sample \citep{Torres2019}. In particular, we have been able to disentangle the white dwarf populations from the different components of the Galaxy, i.e., thin and thick disk, and halo. The artificial intelligent method used consisted in a supervised method based on Random Forest techniques.  This Random Forest algorithm was applied to an 8-dimensional space formed by equatorial coordinates, parallax, proper motion components and photometric magnitudes. 
This 8-dimensional space permits the algorithm to maximize the information in order to classify its different components. Our results showed that the algorithm presents an accuracy of 85.3\%. In particular, our analysis indicates that $80\%$ of possible halo white dwarfs in the 100\,pc sample are expected to be correctly identify, and only a low $5\%$ contamination is expected. These scores are higher than any other obtained by tangential velocity cuts, Toomre diagram criteria  or reduced proper motion criteria  usually applied for selecting the different kinematic populations. Moreover, practically all white dwarfs within 100\,pc that were previously identified in the literature as halo members  are also included in our sample. Thus, the sample found represents the most complete and largest volume-limited sample of the halo white dwarf population to date. However, these facts do not exclude the need to perform a detailed analysis of the completeness of the sample, in particular, for the faint region of the sample. This is done in Section \ref{ss:comple}.

The sample consists of 95 halo white dwarfs candidates representing an old and high velocity population. Its space density is $(4.8\pm0.4)\times10^{-5}\,{\rm pc^{-3}}$, accounting for $1\%$ of the whole white dwarf population within 100\,pc. A complete list of the 95 halo members with their {\it Gaia} source ID, coordinates and main parameters is shown in Table 3 of \cite{Torres2019}. In Figure \ref{f:HR-tracks} we plot our 95 halo candidates in the {\it Gaia} Hertzsprung-Russell (HR) diagram. For illustrative purposes we also plot the cooling tracks belonging to hydrogen-rich and pure helium atmosphere models for different masses. As shown in Fig. \ref{f:HR-tracks}, many objects lie near the blue-hook of the hydrogen-rich atmosphere tracks, thus representing cold and old objects. At first sight, however, it seems to reveal that many of these objects have low masses, $\lappr0.5\,M_{\odot}$, even when compared with He-pure atmosphere tracks. This effect was analyzed by  \cite{Bergeron2019} claiming that probably most of the  possible non-DA objects in this region are DA or a large fraction of unresolved double degenerates populates this region. However, we shall see in Section \ref{ss:syn} that this apparent shift towards lower masses is nicely resolved when taking into account photometric and astrometric errors in the observed data.

\subsection{Spectroscopic observation of halo white dwarfs}
\label{ss:spe}

\begin{figure}
      {\includegraphics[width=\columnwidth]{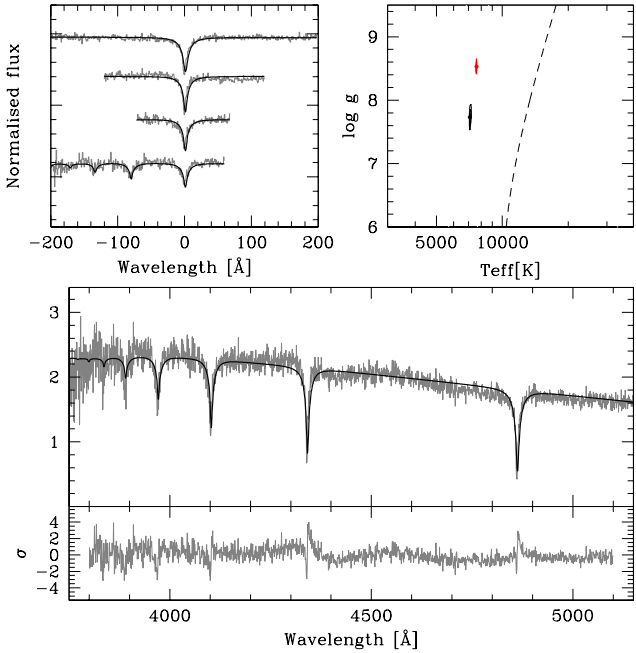}}
  \caption{Spectral model fit to the DA white dwarf J0148--1712. Top left panel: best-fit (black lines) to the normalised Balmer line profiles sampled by the FORS2 spectrum (gray  lines). Top right panel: 3, 5, and 10$\sigma$  $\chi^2$ contour plots in the T$_\mathrm{eff}-\log g$ plane. The black contours refer to the best line profile fit, the red contours to the fit of the whole spectrum. The dashed line  indicates the occurrence of maximum H$\beta$  equivalent width. The best ``hot'' (not visible in the figure) and ``cold'' line profile solutions are indicated by black dots, the  best fit to the whole spectrum is indicated by a red dot. Bottom panels: the white dwarf spectrum (gray line) along with the best-fit  white dwarf model (black line) (top)  and the  residuals of the fit (gray line, bottom).} 
  \label{f:DAfit}
\end{figure}

\begin{figure}
      {\includegraphics[width=\columnwidth]{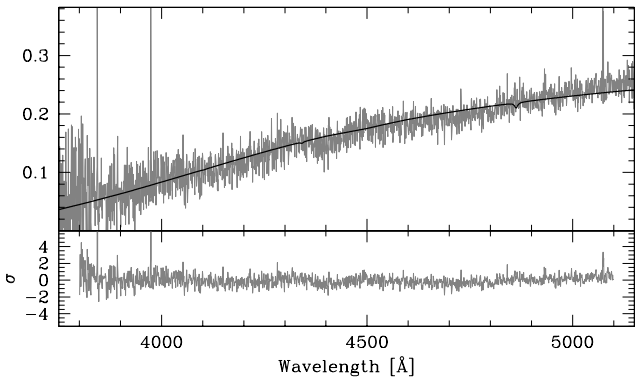}}
  \caption{Spectral model fit to the DC white dwarf J2117--4156. Top panel: the white dwarf spectrum (gray line) along with the best-fit  white dwarf model (black line). Bottom panel: the  residuals of the fit (gray line).} 
  \label{f:DCfit}
\end{figure}

We obtained low-resolution spectra for 27 of our 95 halo white dwarf candidates with the Very Large Telescope UT1 equipped with the FOcal Reducer/low-dispersion Spectrograph (FORS2) \citep{Appenzeller1998}. We used the GRIS\_1200+97 grism and the 1" slit width, which resulted in spectra covering the $\simeq$3800-5200\,\AA\, wavelength range at a resolving power of $R\simeq$1400. We used the {\sc{Pamela}} software \citep{Marsh1989} to subtract the sky contribution and to extract the one dimensional spectra. The data were then wavelength and flux calibrated using arc lamps and flux standard stars taken at the same time of the observations within the {\sc{molly}} package\footnote{Tom Marsh’s {\sc{molly}} package is available at http://deneb.astro.warwick.ac.uk/phsaap/software}. Visual inspection of the flux-calibrated spectra revealed that 24 targets were DC white dwarfs, a result which is not surprising given that these objects are expected to be very old and cold. The remaining three white dwarfs were classified as two DQs (J0248-3001 and J1159-4629) and one DA (J0148-1712). The {\it Gaia} source ID, short name, {\it Gaia} absolute magnitude $M_{\rm G}$, colour $G_{\rm{\tiny BP}}\!-\!G_{\rm{\tiny RP}}$ and spectral type of the 24 DC+1 DA observed white dwarfs are, respectively, provided in the first five columns of Table \ref{t:spec}.

In order to derive the stellar parameters of the white dwarfs with available spectra we used the fitting routine of \citet{Rebassa2007}. Briefly, this procedure uses a grid of model spectra \citep{Koester10}\footnote{The grid includes 1260 spectra of effective temperatures between 3000 and 40,000\,K in steps of 250\,K up to 20,000\,K, steps of 1000\,K up to 30,000\,K and steps of 2000\,K up to 40,000\,K; and surface gravities between 6 and 9.5\,dex in steps of 0.25\,dex for each effective temperature. Prior to the fit, the model spectra were folded at the resolving power of the observed spectra.} to fit both the continuum and the normalised Balmer lines of the spectra to derive the effective temperatures and surface gravities. By adopting then a white dwarf mass-radius relation \citep[e.g.][]{Camisassa2016}, the mass and radius (hence bolometric luminosity since the temperature is known) are also obtained. This routine is only valid if the white dwarfs are hydrogen-rich DAs. In this case, the fit to the continuum is only used to differentiate between the so-called hot and cold solutions obtained from the Balmer line fitting. The fit to our only DA halo white dwarf candidate with available FORS2 spectrum is shown in Figure\ref{f:DAfit}. However, for our DC white dwarfs, which are the vast majority of the spectroscopically observed sample, we could only rely on the fits to the continuum since these objects are absent of any lines. Thus, only the effective temperatures were obtained in these cases. A fit to one of our DC white dwarf halo candidates can be seen in Figure\ref{f:DCfit}.

We note that another possibility exists for deriving the white dwarf radii of our DC white dwarfs by using the flux scaling factors between the observed spectra and the best fit models, since the distances are known from the {\it Gaia} parallaxes. Hence, a spectroscopic bolometric luminosity could be obtained from the effective temperatures and radii via the Stefan-Boltzmann equation. We performed a first attempt to follow this procedure by first re-scaling the observed spectra to the {\it Gaia} fluxes (and PanStarrs fluxes when available too) of our stars. 

The radii, effective temperatures and bolometric luminosities thus derived for our DA and the rest of DC  halo white dwarf candidates are shown, respectively, in the last three columns of Table \ref{t:spec}. 
In Section \ref{ss:methtest} we discuss the validity of the spectroscopic parameter values thus obtained.

\begin{table*}
\caption{Spectroscopic determinations of the parameters of the halo white dwarf candidates for which the spectrum is available. The stellar parameters derived, i.e., radius, effective temperature and bolometric luminosity, are shown, respectively, in the last three columns.}
\label{t:spec}
\begin{center}
\begin{tabular}{cccccccc}  
\hline \hline 
Gaia	             &  Short 	 & $M_{\rm G}$ &  $G_{\rm{\tiny BP}}\!-\!G_{\rm{\tiny RP}}$ &  Spectral & 	Radius &  $T_{\rm eff}$ &  Luminosity \\
source ID			     &  name          & (mag) & (mag) & type &    $(R_{\odot}/10^3)$ &  $(K)$    & $(L_{\odot}/10^5)$ \\  
			     \hline 
5042228731477861888 &	J0129-2257 & 15.55 &  1.41 & DC & $11.76\pm 0.97$&	$4094\pm100$ & $3.49\pm 0.83$ \\
5142197118950177280 &	J0148-1712 & 13.12 & 0.47 & DA &	$15.90	\pm 1.09$&	$7144\pm50$ & $59.15\pm10.56$ \\
2490975272405858048 &	J0205-0517 &  15.59 & 1.53 &  DC &	$10.64\pm 0.69$&	$4094\pm100$ & $2.86\pm0.58$ \\
4616895783694397184 & 	J0237-8445 & 15.77 &  1.57 &   DC & 	$13.25\pm 2.31$&	$3955\pm100$ & $3.85\pm1.64$ \\
5188044687948351872 &	J0301-0044 & 15.30 &  1.55 &  DC &  	$10.89\pm 1.05$&	$4287\pm100$ & $3.60\pm0.95$ \\
4862884499360563968 &	J0340-3301 & 15.79 & 1.32 &  DC & 	$10.58\pm 0.89$&	$4238\pm100$ & $3.24\pm0.77$ \\
3249657094642979840 &	J0342-0344 & 15.50 &  1.11 &   DC &  	$10.90\pm 0.91$&	$4541\pm100$ & $4.54\pm1.05$ \\
2989049057626796416 &	J0518-1155 & 15.70 & 1.74&   DC &  	$9.45\pm 1.02$ & $4001\pm	100$ & $2.05\pm0.60$ \\
5228861484450843648 &	J1049-7400 & 16.09 & 1.55 &  DC &	$8.22\pm 0.87$& $4047\pm	100$ & $1.63\pm0.47$ \\
3801499128765222400 &	J1053-0307 & 15.39 & 1.44  & DC &   	$11.19\pm 0.98$ & $4141\pm	100$ & $3.31\pm0.82$ \\
5348874243767794304 &	J1123-5150 & 15.61 & 1.34 &  DC &  	$11.89\pm 1.19$ & $4865\pm	100$ & $2.83\pm1.87$ \\
5377861317357370240 &	J1159-4630 & 15.19 & 0.42 &   DC &  	$8.32\pm 0.74$ & $4489\pm	100$ & $2.52\pm0.62$ \\
6085402414245451520 &	J1312-4728 & 15.58 &  1.54 &  DC & 	$10.58\pm 0.79$ & $4094\pm	100$ & $2.83\pm0.63$ \\
6165095738576250624 &	J1342-3415 & 14.66 &  1.05 &   DC &   	$10.88\pm 0.85$ & $5095\pm	100$ & $7.16\pm1.54$ \\
5824436284328653312 &	J1517-6645 & 16.01 &  1.31 &  DC & 	$9.71\pm 1.33$ & $3865\pm	100$ & $1.89\pm0.67$ \\
6007140379167609984 &	J1518-3803 & 15.68 & 1.56 &  DC &  	$10.51\pm1.11$ & $4189\pm	100$ & $3.05\pm0.87$ \\
5827557213731539328 &	J1539-6124 & 15.23 & 1.33 &   DC &   	$11.62\pm0.89$& $4336\pm	100$ & $4.29\pm0.96$ \\
5817295536128445568 &	J1707-6319 & 15.12 & 1.41 &  DC &   	$11.25\pm0.94$& $4437\pm	100$ & $4.40\pm1.04$ \\
6647162730439433984 &	J1936-4913 & 15.60 &  1.51 &   DC & 	$10.77\pm0.99$& $4189\pm	100$ & $3.21\pm0.82$ \\
6471523921227261056 &	J2042-5218 & 14.70 & 0.94 &  DC  & 	$11.65\pm0.86$& $4755\pm	100$ & $6.23\pm1.31$ \\
6580458035746362496 &	J2117-4156 & 15.16 & 1.34 &  DC & 	$11.80\pm0.84$& $4437\pm	100$ & $4.85\pm1.02$ \\
6580551872194787968 &	J2129-0034 & 15.39 & 1.58 &  DC & 	$12.60\pm1.53$& $4287\pm	100$ & $4.81\pm1.51$ \\
2687584757658775424 &	J2230-7515 & 15.58  & 1.55  & DC  & 	$10.62\pm0.82$ & $4141\pm	100$ & $2.98\pm0.67$ \\
6357629089412187648 &	J2319-0613 & 15.07 & 1.38 &   DC  &  	$7.05\pm0.74$ & $4979\pm	100$ & $2.75\pm0.74$ \\
2631967439437024384 &	J2349-0124 & 15.01 & 1.23 &  DC & 	$13.41\pm1.08$ & $4287\pm	100$ & $5.45\pm1.26$ \\

\hline \hline
\end{tabular}
\end{center}
\end{table*}

\subsection{The synthetic halo white dwarf population}
\label{ss:syn}

We complement our analysis of the 95 halo white dwarfs candidates with the aid of a detailed population synthesis code. Our code, based on Monte Carlo techniques, has been widely used in  the study of the white dwarf population of the different Galactic components, i.e., disk, halo and bulge,  as well as in globular and open clusters \citep[e.g.][]{GBerro1999, GBerro2004, GBerro2010, Torres2001, Torres2002, Torres2015, Torres2016, Torres2018}. Here we will only mention the main ingredients of our halo simulation, while a thorough description of the code, as well as its {\it Gaia} performances, can be found in \cite{Torres2005,Jimenez2018,Torres2019}.

The main objective of our halo white dwarf simulation is to over-populate the entire white dwarf region of the HR-diagram space. Synthetic white dwarfs will be used then to extract the physical parameters at the HR-diagram loci where observed white dwarfs have been found. For this reason, it is not necessary to take into account a meaningful star formation history, nor an exact age of the halo population. Consequently, we just adopt a constant star formation rate and we generate stars with an upper-limiting age fixed at the age of the Universe of 13.7\,Gyr \citep{PlanckCol2016}. Stars generated are drawn form a Salpeter initial mass function with a standard slope value of $\alpha=-2.35$. We considered that white dwarfs are formed only through single evolution. Additionally,  as done in the halo white dwarf analysis by \cite{Kilic2018}, a constant  metallicity value of $[{\rm Fe/H}]=-1.5$ is adopted. Main-sequence lifetimes are drawn from BaSTI models \citep{Hidalgo2018} and the semi-empirical initial-to-final mass relationship of \cite{Catalan2008} is applied. White dwarf cooling times are derived from a complete set of cooling sequences,  which encompass the full range of masses \citep{Althaus2015,Camisassa2017,Camisassa2019}. It is worth noting that these cooling sequences are specifically calculated for the adopted metallicity and are the result of the full previous progenitor evolution, starting at the zero-age main sequence, all the way through central hydrogen and helium burning, thermally pulsing asymptotic giant branch (AGB) and post-AGB phases. The lifetime of the white dwarf progenitors along with the initial-to-final mass relationship obtained this way are in completely agreement with those input models previously stated. This fact guarantees us the use of a coherent set of evolutionary sequences.
Hydrogen-rich and hydrogen-deficient atmosphere models are generated according to the canonical distribution of $80\%$  and $20\%$, respectively. For the spatial distribution an isothermal model is adopted, which is practically equivalent to an isotropic distribution for the local 100\,pc neighborhood. Magnitudes are interpolated in the corresponding cooling sequences and calculated in the {\it Gaia} filters  (R. Rohrmann's private communication) using the appropriate atmosphere models. Finally, photometric and astrometric errors are added following {\it Gaia}'s performance\footnote{http://www.cosmos.esa.int/web/gaia/science-performance}.

A representative HR-diagram of our simulated halo white dwarf sample is presented in the right panel of Fig. \ref{f:HR-tracks}. The number of objects generated is large enough to ensure statistical significance in all the regions around the observed objects. As it  can be seen in Fig. \ref{f:HR-tracks}, the synthetic sample populates all the regions where our 95 halo white dwarf candidates are located. Only one object (J0055+3847) is outside of the simulated space, and this is probably because it is either a He-core white dwarf or a double degenerate system. It is also worth noting that those objects previously discussed in Section \ref{ss:ide} as cold and abnormally shifted in the HR-diagram towards very low masses are now naturally recovered in the simulated space, once photometric and astrometric errors are taken into account.

\begin{figure*}
      {\includegraphics[width=1.0\columnwidth, clip=true,trim=5 20 15 30]{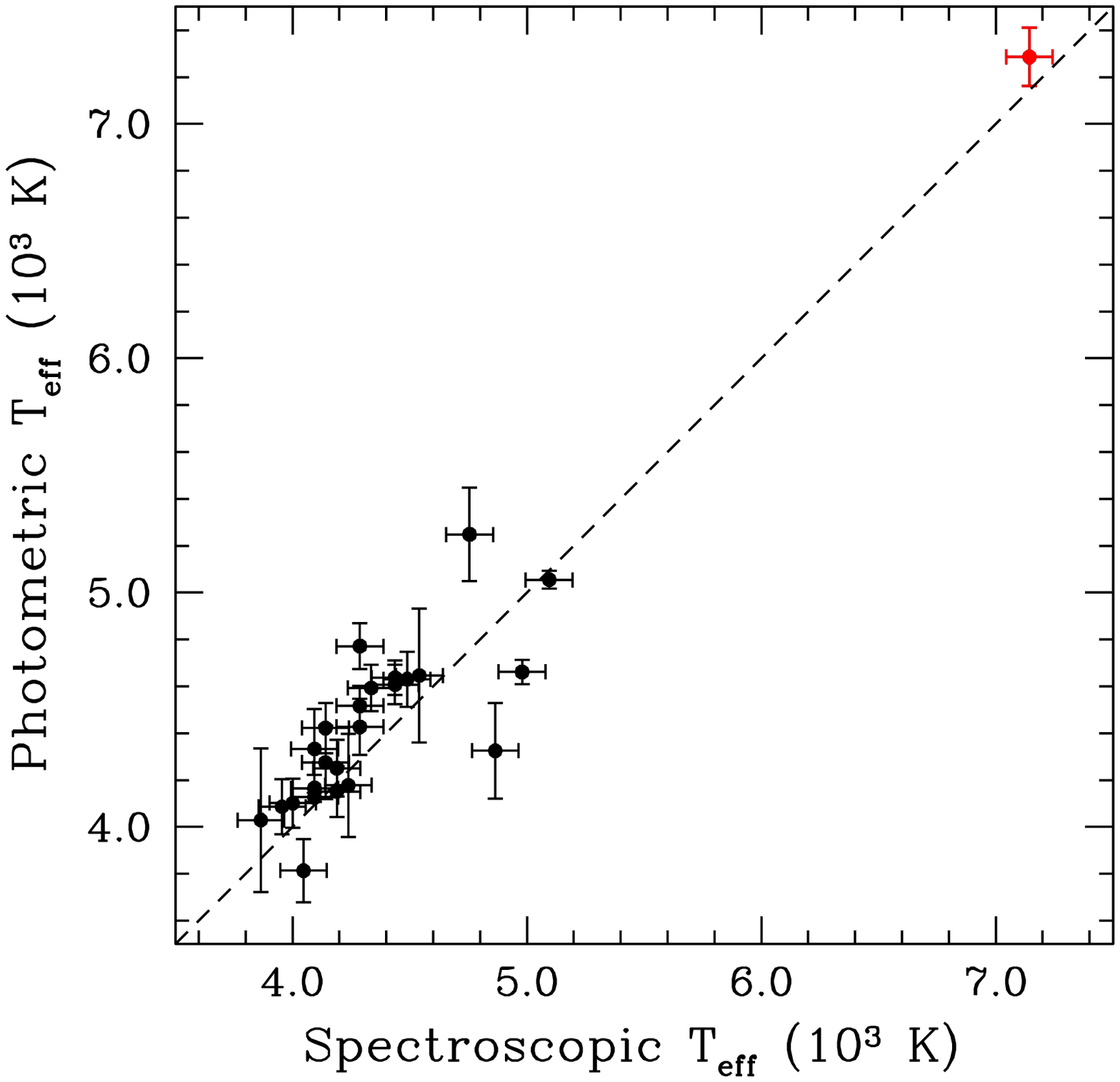}}
            {\includegraphics[width=1.0\columnwidth, clip=true,trim=5 20 15 30]{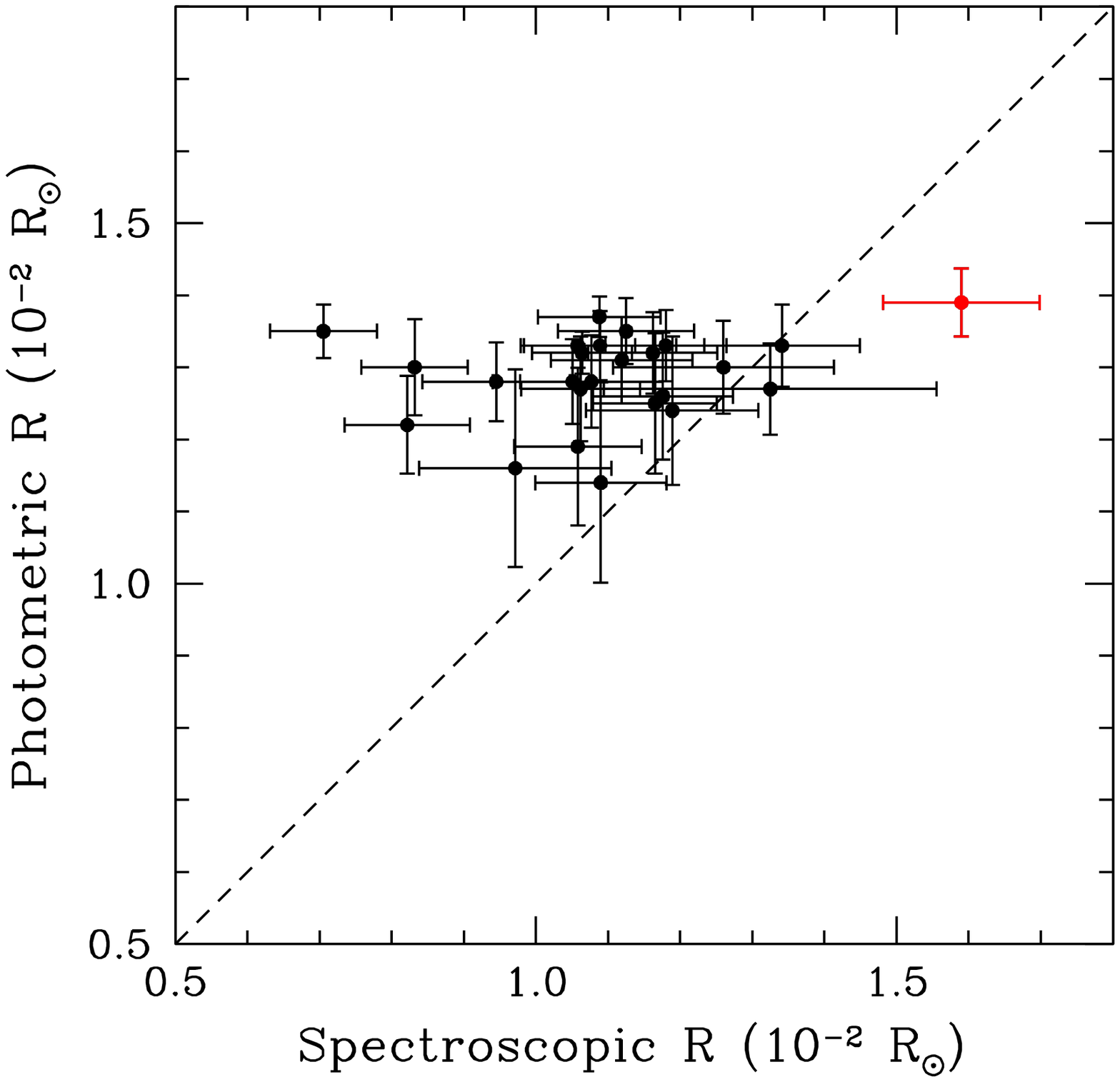}}
      {\includegraphics[width=1.0\columnwidth, clip=true,trim=5 20 15 30]{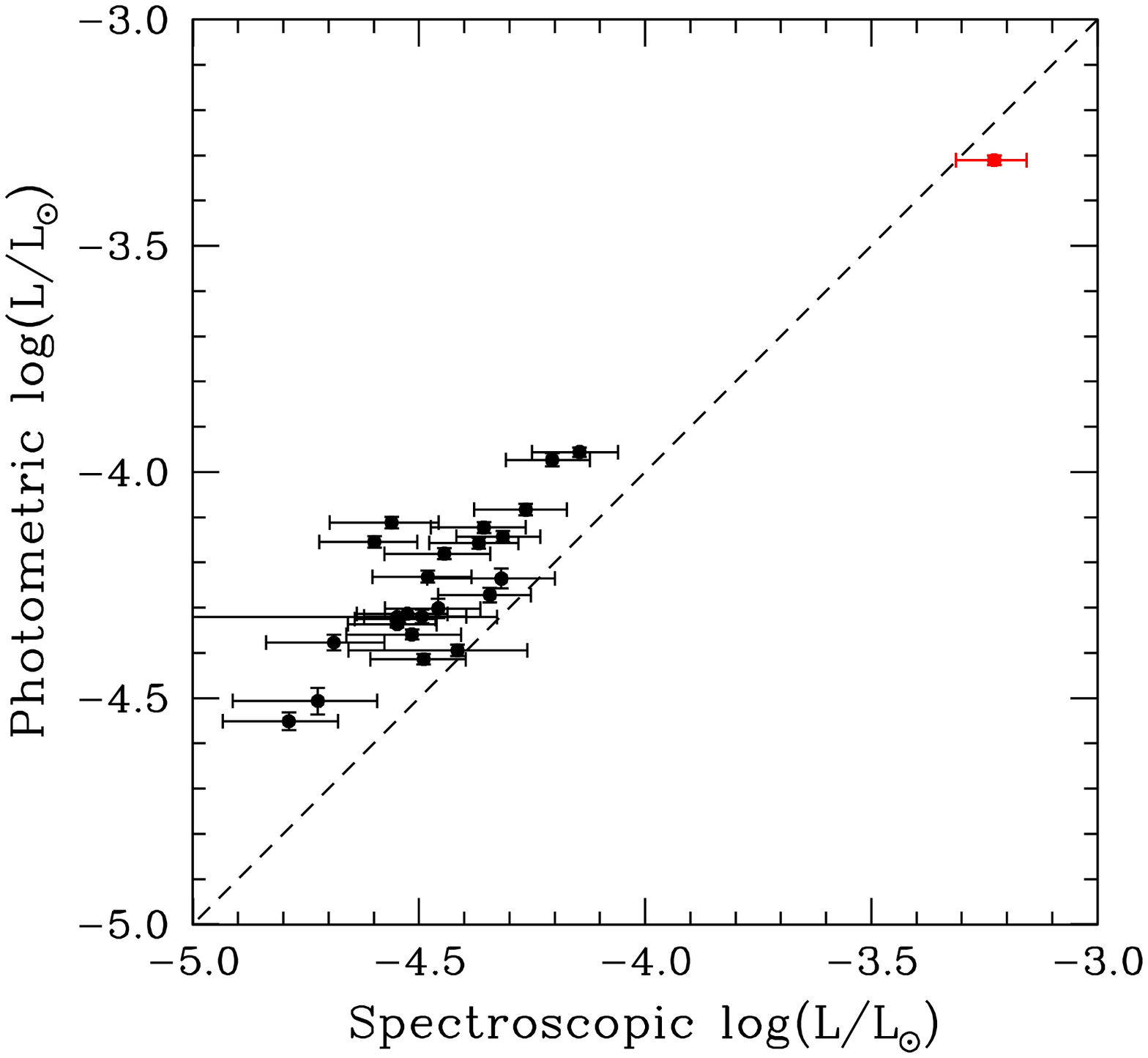}}
            {\includegraphics[width=1.0\columnwidth, clip=true,trim=5 20 15 30]{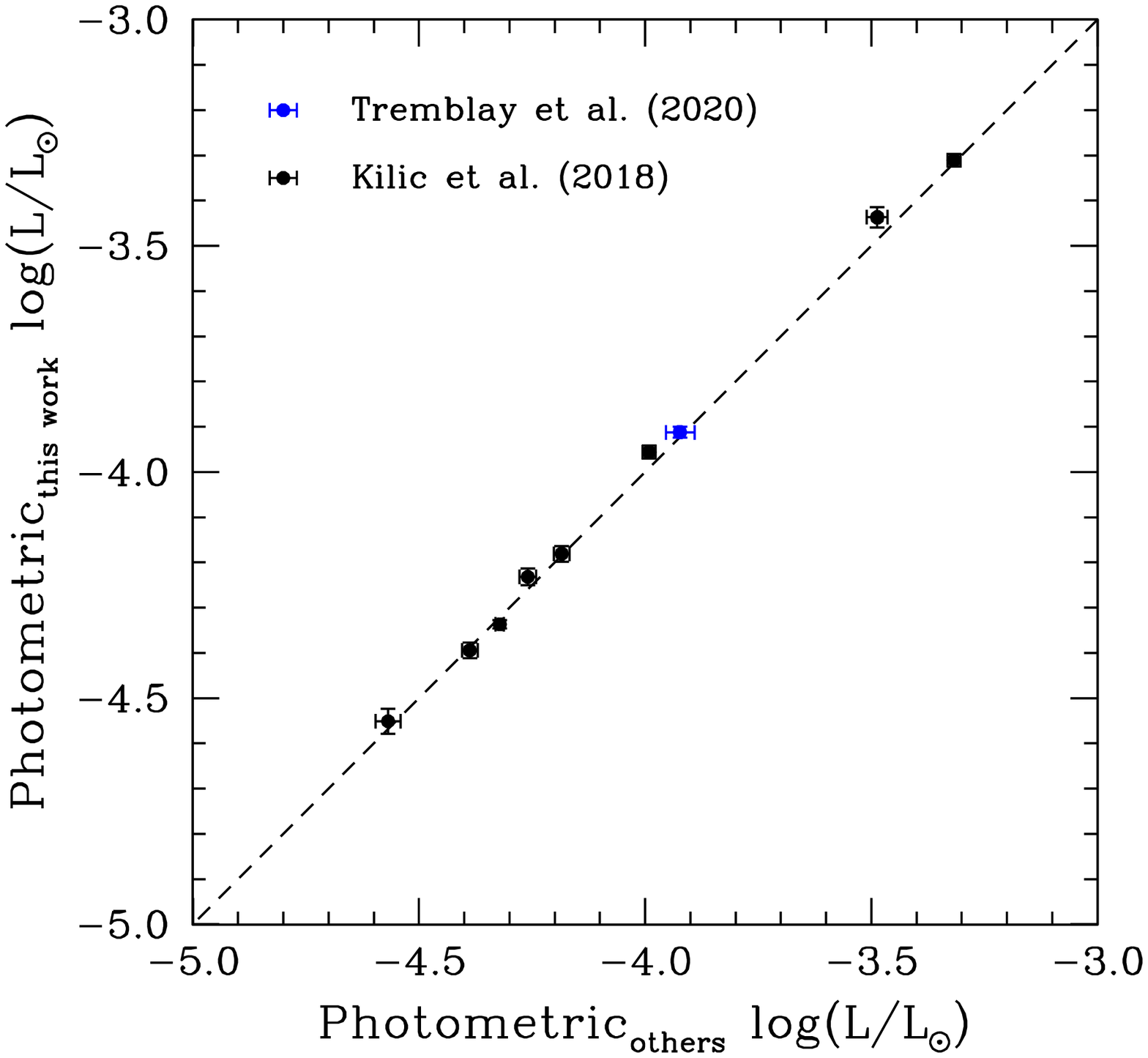}}
  \caption{Spectroscopic effective temperatures (top left panel), radii (top right panel) and luminosities (bottom left panel) for the 24 DC (black dots) and one DA (red dot) observed objects (see Section \ref{ss:spe}) compared to the respective photometric  parameters derived as the average value within $1\sigma$ ellipses.  Also shown (bottom right panel) is the photometric luminosity obtained in this work with other estimations also photometrically obtained. The equality line is plotted in all panels as a dashed line. The photometric effective temperatures agree with those spectroscopically derived. That is not the case of the radii, where intrinsic errors when applying  flux calibration induce a systematic error in the radius calculation in the spectroscopic method, hence in the bolometric luminosities also. On the contrary, the luminosities derived by our photometric procedure are in perfect agreement with those published in the literature (see text for details).} 
  \label{f:lum}
\end{figure*}

\section{Results}

\subsection{Methodology and testing}
\label{ss:methtest}

The population synthesis sample described in the previous section allows us to derive the representative white dwarf parameters at each locus of the HR-diagram by just taking into account the {\it Gaia} photometric and astrometric errors. For each of the 95 objects of our halo sample we first derive the theoretical errors in $G$ magnitude and $G_{\rm BP}-G_{\rm RP}$ colour, $\sigma_G$ and $\sigma_{G_{\rm BP}-G_{\rm RP}}$, respectively. We build then a region centered in the location of each object in the HR-diagram that contains all the synthetic stars within $1\sigma$. The ellipses thus formed are shown in the right panel of Fig. \ref{f:HR-tracks}. For the sake of clarity we only plot some of the most representatives. The $G$-magnitude error is mainly due to the photometric error in the $G$ filter, given that the astrometric error in parallax is practically negligible. On the contrary, the error in  $G_{\rm BP}-G_{\rm RP}$ colour is the addition of the photometric errors from $G_{\rm BP}$ and $G_{\rm RP}$ filters. Thus, in general terms, the size of the $1\sigma$-ellipses increases for dimmer objects, being the error in the absolute $G$ magnitude substantially smaller than the error in colour.

For each of the $1\sigma$ ellipses we obtain the average luminosity,  mass, age, effective temperature and radius of all synthetic white dwarfs within it. We can now compare the photometric parameters thus obtained with the spectroscopic ones derived for the 25 observed objects presented in Section \ref{ss:spe}. The results are plotted in Figure \ref{f:lum}. On the top panels we show the comparison between the photometric versus the spectroscopic effective temperature (top left panel) and white dwarf radius (top right panel). The agreement in effective temperatures is good, since most of the objects are within the $1\sigma$ errors. However, that is not the case for the white dwarf radii. Those obtained from the spectroscopic analysis are systematically lower than those obtained photometrically. This effect is most likely due to intrinsic errors when applying the flux calibration to our FORS2 spectra, which directly affect the radii measurements through the flux scaling factors.

Consequently, the luminosity thus obtained from the spectroscopic method suffers from this imprecision. In the bottom left panel of Fig. \ref{f:lum} we compare the photometric versus the spectroscopic luminosity for the objects in which the spectrum is available. Photometric luminosities are shifted towards larger (brighter) luminosities than those derived spectroscopically. That is a consequence of systematically lower spectroscopic radii, as previously stated.  For this reason, we do not consider the spectroscopic radii (hence bolometric luminosities) to be as reliable as the photometric ones and hence decided not to make use of these values.

Finally, we carry out an additional verification comparing the  photometric luminosity of our work with other values  published in the literature and also derived photometrically \citep{Kilic2018,Tremblay2020}. The results are shown in the bottom right panel of Fig. \ref{f:lum}. As it can be seen, the agreement is excellent.  Thus, our photometric procedure provides effective temperatures which are in agreement with those derived spectroscopically and yields luminosities which are in perfect agreement with those also derived  photometrically in the literature. Consequently, we can conclude that the rest of parameters are  robustly obtained and derived in a consistent way by our photometric method.

Once we have demonstrated that our strategy is feasible to compute  the white dwarf luminosities we extend it to the whole halo white dwarf sample. Analogously, we also derive the rest of parameters: mass, age, effective temperature and radius. The only exception is for object J0055+3847, which is located, as previously stated, outside the region of the HR-diagram covered by the single white dwarf population (see Fig. \ref{f:HR-tracks}). In Table \ref{t:photo} we present the  parameters thus derived from our photometric method. Along with the {\it Gaia} source ID (first column) and the short name (second column), we show the absolute magnitude and colour and their respective errors used in building our ellipsoids (columns third to sixth) and the mass, effective temperature, bolometric luminosity and age, with their respective errors (columns seventh to tenth). It is worth recalling that our photometric method assumes a mixed population of $80\%\,$DA and $20\%\,$DB white dwarfs, and then just counts the physical parameters not caring what the atmospheric composition is.

\begin{table*}
\caption{Stellar parameters derived by our photometric method for the whole halo white dwarf sample within 100\,pc.
Along with the {\it Gaia} Source ID and short name (first and second columns, respectively), the location within the HR-diagram (third and fourth columns) and the size of the adopted ellipses (fifth and six columns) are shown. 
Radius, effective temperature,   bolometric luminosity and total age are shown, respectively, in the last four columns.}
\label{t:photo}
\begin{center}
\begin{tabular}{cccccccccc}  
\hline \hline 
Gaia & Short	 & $M_{\rm G}$ &  $G_{\rm{\tiny BP}}\!-\!G_{\rm{\tiny RP}}$ & $\sigma_{M_{\rm G}}$ &  $\sigma_{G_{\rm{\tiny BP}}\!-\!G_{\rm{\tiny RP}}}$ &  Mass & $T_{\rm eff}$ &	Luminosity & Age  \\
Source ID & name     & (mag) & (mag) & (mag) & (mag) & $M_{\odot}$ & (K)  & $\log(L/L_{\odot})$ & (Gyr) \\  
			     \hline 
  420531621029108608	&	 J0013+5438	 	&	 	15.47	 	&	 	1.47	 	&	 	0.007	 	&	 	0.041	 	&	 	$	0.551	\pm	0.027	$ 	&	 $	4306	\pm	73	$ 	&	 $	-4.273	\pm	0.007	$ 	&	 $	11.27	\pm	1.83	$	\\	 
  5006232026455470848	&	 J0045-3329	 	&	 	16.05	 	&	 	1.51	 	&	 	0.011	 	&	 	0.072	 	&	 	$	0.613	\pm	0.051	$ 	&	 $	3829	\pm	128	$ 	&	 $	-4.544	\pm	0.019	$ 	&	 $	12.22	\pm	1.48	$	\\	 
  367799116372410752	&	 J0055+3847	 	&	 	13.51	 	&	 	0.89	 	&	 	0.015	 	&	 	0.037	 	&	 	$\cdots$ 	&	$\cdots$	&	$\cdots$	&	$\cdots$\\
  2583365245917474816	&	 J0106+1141	 	&	 	15.76	 	&	 	1.43	 	&	 	0.039	 	&	 	0.169	 	&	 	$	0.599	\pm	0.065	$ 	&	 $	4112	\pm	163	$ 	&	 $	-4.404	\pm	0.013	$ 	&	 $	11.43	\pm	1.39	$	\\	 
  5042228731477861888	&	 J0129-2257	 	&	 	15.54	 	&	 	1.41	 	&	 	0.086	 	&	 	0.282	 	&	 	$	0.589	\pm	0.067	$ 	&	 $	4333	\pm	170	$ 	&	 $	-4.302	\pm	0.022	$ 	&	 $	10.56	\pm	1.56	$	\\	 
  96095735719745280	&	 J0132+1941	 	&	 	14.85	 	&	 	1.13	 	&	 	0.049	 	&	 	0.145	 	&	 	$	0.556	\pm	0.049	$ 	&	 $	4941	\pm	130	$ 	&	 $	-4.033	\pm	0.015	$ 	&	 $	8.36	\pm	2.41	$	\\	 
  5142197118950177280	&	 J0148-1712	 	&	 	13.11	 	&	 	0.46	 	&	 	0.011	 	&	 	0.025	 	&	 	$	0.529	\pm	0.023	$ 	&	 $	7286	\pm	124	$ 	&	 $	-3.311	\pm	0.010	$ 	&	 $	6.52	\pm	2.66	$	\\	 
  343356212681211392	&	 J0157+3932	 	&	 	15.64	 	&	 	1.44	 	&	 	0.046	 	&	 	0.187	 	&	 	$	0.585	\pm	0.054	$ 	&	 $	4205	\pm	135	$ 	&	 $	-4.350	\pm	0.013	$ 	&	 $	11.03	\pm	1.44	$	\\	 
  521409549427243904	&	 J0158+6904	 	&	 	15.47	 	&	 	1.14	 	&	 	0.103	 	&	 	0.298	 	&	 	$	0.630	\pm	0.108	$ 	&	 $	4516	\pm	279	$ 	&	 $	-4.272	\pm	0.024	$ 	&	 $	9.80	\pm	1.60	$	\\	 
  2490975272405858048	&	 J0205-0517	 	&	 	15.58	 	&	 	1.52	 	&	 	0.008	 	&	 	0.046	 	&	 	$	0.544	\pm	0.019	$ 	&	 $	4165	\pm	57	$ 	&	 $	-4.325	\pm	0.009	$ 	&	 $	11.81	\pm	1.42	$	\\	 
  4616895783694397184	&	 J0237-8445	 	&	 	15.74	 	&	 	1.55	 	&	 	0.039	 	&	 	0.176	 	&	 	$	0.579	\pm	0.047	$ 	&	 $	4086	\pm	117	$ 	&	 $	-4.394	\pm	0.012	$ 	&	 $	11.62	\pm	1.38	$	\\	 
  5065611697758431360	&	 J0248-3001	 	&	 	13.91	 	&	 	0.71	 	&	 	0.019	 	&	 	0.050	 	&	 	$	0.573	\pm	0.040	$ 	&	 $	6143	\pm	162	$ 	&	 $	-3.669	\pm	0.011	$ 	&	 $	4.85	\pm	2.02	$	\\	 
  5188044687948351872	&	 J0301-0044	 	&	 	15.24	 	&	 	1.51	 	&	 	0.041	 	&	 	0.157	 	&	 	$	0.543	\pm	0.033	$ 	&	 $	4516	\pm	86	$ 	&	 $	-4.181	\pm	0.012	$ 	&	 $	10.60	\pm	1.79	$	\\	 
  4862884499360563968	&	 J0340-3301	 	&	 	15.79	 	&	 	1.31	 	&	 	0.032	 	&	 	0.143	 	&	 	$	0.636	\pm	0.087	$ 	&	 $	4177	\pm	220	$ 	&	 $	-4.414	\pm	0.012	$ 	&	 $	11.17	\pm	1.23	$	\\	 
  3249657094642979840	&	 J0342-0344	 	&	 	15.46	 	&	 	1.09	 	&	 	0.057	 	&	 	0.194	 	&	 	$	0.680	\pm	0.112	$ 	&	 $	4646	\pm	285	$ 	&	 $	-4.272	\pm	0.016	$ 	&	 $	9.21	\pm	1.28	$	\\	 
  4857106909354185344	&	 J0345-3611	 	&	 	15.37	 	&	 	1.47	 	&	 	0.061	 	&	 	0.212	 	&	 	$	0.559	\pm	0.043	$ 	&	 $	4422	\pm	110	$ 	&	 $	-4.235	\pm	0.015	$ 	&	 $	10.54	\pm	1.74	$	\\	 
  66837563803594880	&	 J0346+2455	 	&	 	15.58	 	&	 	1.50	 	&	 	0.013	 	&	 	0.069	 	&	 	$	0.553	\pm	0.031	$ 	&	 $	4197	\pm	83	$ 	&	 $	-4.322	\pm	0.009	$ 	&	 $	11.49	\pm	1.51	$	\\	 
  4864861112027378944	&	 J0431-3816	 	&	 	15.75	 	&	 	1.77	 	&	 	0.133	 	&	 	0.456	 	&	 	$	0.584	\pm	0.053	$ 	&	 $	4089	\pm	143	$ 	&	 $	-4.399	\pm	0.029	$ 	&	 $	11.54	\pm	1.37	$	\\	 
  4864752883148064512	&	 J0432-3902	 	&	 	14.86	 	&	 	1.25	 	&	 	0.005	 	&	 	0.025	 	&	 	$	0.527	\pm	0.021	$ 	&	 $	4846	\pm	57	$ 	&	 $	-4.037	\pm	0.013	$ 	&	 $	9.86	\pm	2.08	$	\\	 
  2989049057626796416	&	 J0518-1155	 	&	 	15.70	 	&	 	1.74	 	&	 	0.079	 	&	 	0.306	 	&	 	$	0.568	\pm	0.040	$ 	&	 $	4102	\pm	104	$ 	&	 $	-4.377	\pm	0.017	$ 	&	 $	11.69	\pm	1.32	$	\\	 
  192454873200555392	&	 J0559+4248		&	 	13.42	 	&	 	0.52	 	&	 	0.022	 	&	 	0.049	 	&	 	$	0.578	\pm	0.035	$ 	&	 $	6992	\pm	172	$ 	&	 $	-3.445	\pm	0.010	$ 	&	 $	3.71	\pm	1.60	$	\\	 
  977441274176008192	&	 J0711+4607	 	&	 	14.76	 	&	 	1.16	 	&	 	0.033	 	&	 	0.105	 	&	 	$	0.536	\pm	0.034	$ 	&	 $	4983	\pm	93	$ 	&	 $	-3.996	\pm	0.014	$ 	&	 $	9.03	\pm	2.42	$	\\	 
  5613373001468333696	&	 J0732-2558	 	&	 	15.11	 	&	 	1.27	 	&	 	0.097	 	&	 	0.265	 	&	 	$	0.568	\pm	0.059	$ 	&	 $	4714	\pm	155	$ 	&	 $	-4.130	\pm	0.023	$ 	&	 $	9.20	\pm	2.14	$	\\	 
  874900643675606912	&	 J0745+2626	 	&	 	15.76	 	&	 	1.60	 	&	 	0.014	 	&	 	0.081	 	&	 	$	0.572	\pm	0.046	$ 	&	 $	4046	\pm	112	$ 	&	 $	-4.406	\pm	0.012	$ 	&	 $	11.80	\pm	1.40	$	\\	 
  1110759459929880704	&	 J0748+7141	 	&	 	15.74	 	&	 	1.62	 	&	 	0.062	 	&	 	0.251	 	&	 	$	0.579	\pm	0.048	$ 	&	 $	4078	\pm	124	$ 	&	 $	-4.398	\pm	0.016	$ 	&	 $	11.67	\pm	1.37	$	\\	 
  3144837318276010624	&	 J0750+0711a	 	&	 	15.07	 	&	 	1.25	 	&	 	0.002	 	&	 	0.013	 	&	 	$	0.530	\pm	0.023	$ 	&	 $	4664	\pm	56	$ 	&	 $	-4.108	\pm	0.006	$ 	&	 $	9.54	\pm	2.07	$	\\	 
  3144837112117580800	&	 J0750+0711b	 	&	 	15.31	 	&	 	1.41	 	&	 	0.002	 	&	 	0.010	 	&	 	$	0.565	\pm	0.025	$ 	&	 $	4512	\pm	59	$ 	&	 $	-4.202	\pm	0.006	$ 	&	 $	10.79	\pm	1.19	$	\\	 
  5726927573083821440	&	 J0822-1249	 	&	 	15.54	 	&	 	1.30	 	&	 	0.037	 	&	 	0.148	 	&	 	$	0.597	\pm	0.069	$ 	&	 $	4346	\pm	170	$ 	&	 $	-4.305	\pm	0.012	$ 	&	 $	10.40	\pm	1.48	$	\\	 
  5742629217603133056	&	 J0912-0953	 	&	 	15.27	 	&	 	1.15	 	&	 	0.094	 	&	 	0.265	 	&	 	$	0.606	\pm	0.090	$ 	&	 $	4656	\pm	230	$ 	&	 $	-4.192	\pm	0.023	$ 	&	 $	9.17	\pm	1.86	$	\\	 
  5215833263797633664	&	 J0913-7553	 	&	 	15.37	 	&	 	1.37	 	&	 	0.038	 	&	 	0.147	 	&	 	$	0.562	\pm	0.046	$ 	&	 $	4439	\pm	116	$ 	&	 $	-4.231	\pm	0.013	$ 	&	 $	10.39	\pm	1.80	$	\\	 
  3840846114438361984	&	 J0925+0018		&	 	15.12	 	&	 	1.15	 	&	 	0.032	 	&	 	0.115	 	&	 	$	0.578	\pm	0.061	$ 	&	 $	4729	\pm	155	$ 	&	 $	-4.134	\pm	0.013	$ 	&	 $	8.83	\pm	2.02	$	\\	 
  1064978578888570496	&	 J0941+6511	 	&	 	15.14	 	&	 	1.34	 	&	 	0.013	 	&	 	0.059	 	&	 	$	0.539	\pm	0.032	$ 	&	 $	4610	\pm	83	$ 	&	 $	-4.139	\pm	0.011	$ 	&	 $	10.49	\pm	2.04	$	\\	 
  3836593100382315904	&	 J1005+0254	 	&	 	15.26	 	&	 	1.29	 	&	 	0.065	 	&	 	0.207	 	&	 	$	0.574	\pm	0.058	$ 	&	 $	4580	\pm	148	$ 	&	 $	-4.187	\pm	0.018	$ 	&	 $	9.62	\pm	1.91	$	\\	 
  746045096445123968	&	 J1012+3233	 	&	 	15.70	 	&	 	1.39	 	&	 	0.131	 	&	 	0.399	 	&	 	$	0.615	\pm	0.091	$ 	&	 $	4224	\pm	240	$ 	&	 $	-4.374	\pm	0.034	$ 	&	 $	11.02	\pm	1.45	$	\\	 
  5192296911732427904	&	 J1036-8225	 	&	 	15.67	 	&	 	1.54	 	&	 	0.040	 	&	 	0.176	 	&	 	$	0.573	\pm	0.044	$ 	&	 $	4150	\pm	109	$ 	&	 $	-4.361	\pm	0.011	$ 	&	 $	11.37	\pm	1.37	$	\\	 
  3862858165427681536	&	 J1036+0732	 	&	 	15.22	 	&	 	1.49	 	&	 	0.023	 	&	 	0.098	 	&	 	$	0.540	\pm	0.027	$ 	&	 $	4530	\pm	75	$ 	&	 $	-4.171	\pm	0.010	$ 	&	 $	10.63	\pm	1.69	$	\\	 
  1076941716370493696	&	 J1036+7110	 	&	 	15.39	 	&	 	1.24	 	&	 	0.002	 	&	 	0.012	 	&	 	$	0.582	\pm	0.071	$ 	&	 $	4461	\pm	165	$ 	&	 $	-4.241	\pm	0.009	$ 	&	 $	10.52	\pm	1.65	$	\\	 
  855361055035055104	&	 J1045+5904	 	&	 	13.89	 	&	 	0.23	 	&	 	0.009	 	&	 	0.029	 	&	 	$	1.074	\pm	0.033	$ 	&	 $	9211	\pm	481	$ 	&	 $	-3.527	\pm	0.024	$ 	&	 $	2.69	\pm	0.25	$	\\	 
  5228861484450843648	&	 J1049-7400	 	&	 	16.07	 	&	 	1.54	 	&	 	0.021	 	&	 	0.122	 	&	 	$	0.613	\pm	0.054	$ 	&	 $	3813	\pm	134	$ 	&	 $	-4.551	\pm	0.020	$ 	&	 $	12.23	\pm	1.54	$	\\	 
  3801499128765222400	&	 J1053-0307	 	&	 	15.37	 	&	 	1.42	 	&	 	0.041	 	&	 	0.156	 	&	 	$	0.556	\pm	0.042	$ 	&	 $	4422	\pm	107	$ 	&	 $	-4.232	\pm	0.013	$ 	&	 $	10.62	\pm	1.77	$	\\	 
  3865951435233552896	&	 J1055+0816	 	&	 	15.46	 	&	 	1.67	 	&	 	0.043	 	&	 	0.181	 	&	 	$	0.556	\pm	0.042	$ 	&	 $	4324	\pm	108	$ 	&	 $	-4.271	\pm	0.014	$ 	&	 $	11.14	\pm	1.62	$	\\	 
  1055533400343235456	&	 J1101+6333	 	&	 	15.33	 	&	 	1.44	 	&	 	0.068	 	&	 	0.226	 	&	 	$	0.560	\pm	0.045	$ 	&	 $	4464	\pm	115	$ 	&	 $	-4.219	\pm	0.017	$ 	&	 $	10.31	\pm	1.81	$	\\	 
  831946229073235200	&	 J1107+4855	 	&	 	15.19	 	&	 	1.27	 	&	 	0.015	 	&	 	0.066	 	&	 	$	0.552	\pm	0.041	$ 	&	 $	4588	\pm	104	$ 	&	 $	-4.162	\pm	0.011	$ 	&	 $	9.97	\pm	1.94	$	\\	 
  5348874243767794304	&	 J1123-5150	 	&	 	15.58	 	&	 	1.32	 	&	 	0.069	 	&	 	0.242	 	&	 	$	0.604	\pm	0.080	$ 	&	 $	4325	\pm	204	$ 	&	 $	-4.321	\pm	0.018	$ 	&	 $	10.57	\pm	1.47	$	\\	 
  856513235846126720	&	 J1123+5742	 	&	 	14.74	 	&	 	0.65	 	&	 	0.061	 	&	 	0.159	 	&	 	$	0.893	\pm	0.100	$ 	&	 $	6163	\pm	418	$ 	&	 $	-3.997	\pm	0.017	$ 	&	 $	5.15	\pm	0.24	$	\\	 
  5224999346778496128	&	 J1147-7457	 	&	 	15.66	 	&	 	1.58	 	&	 	0.003	 	&	 	0.021	 	&	 	$	0.575	\pm	0.037	$ 	&	 $	4159	\pm	95	$ 	&	 $	-4.357	\pm	0.005	$ 	&	 $	11.58	\pm	1.20	$	\\	 
  3892524535332945280	&	 J1151+0159	 	&	 	15.13	 	&	 	1.24	 	&	 	0.061	 	&	 	0.188	 	&	 	$	0.564	\pm	0.054	$ 	&	 $	4683	\pm	139	$ 	&	 $	-4.138	\pm	0.017	$ 	&	 $	9.35	\pm	2.11	$	\\	 
  5377861317357370240	&	 J1159-4630	 	&	 	15.17	 	&	 	1.24	 	&	 	0.029	 	&	 	0.109	 	&	 	$	0.561	\pm	0.046	$ 	&	 $	4630	\pm	118	$ 	&	 $	-4.154	\pm	0.013	$ 	&	 $	9.54	\pm	2.00	$	\\	 
  5377861592235273856	&	 J1159-4629	 	&	 	13.44	 	&	 	0.41	 	&	 	0.009	 	&	 	0.023	 	&	 	$	0.705	\pm	0.051	$ 	&	 $	7589	\pm	241	$ 	&	 $	-3.437	\pm	0.016	$ 	&	 $	2.32	\pm	0.37	$	\\	 
  1573358945589364608	&	 J1205+5502	 	&	 	15.84	 	&	 	1.42	 	&	 	0.106	 	&	 	0.360	 	&	 	$	0.623	\pm	0.089	$ 	&	 $	4091	\pm	233	$ 	&	 $	-4.439	\pm	0.030	$ 	&	 $	11.50	\pm	1.39	$	\\	 
  3905186270720273152	&	 J1217+0830	 	&	 	15.72	 	&	 	1.31	 	&	 	0.075	 	&	 	0.264	 	&	 	$	0.619	\pm	0.089	$ 	&	 $	4219	\pm	227	$ 	&	 $	-4.380	\pm	0.019	$ 	&	 $	10.99	\pm	1.43	$	\\	 
  1533950318546008448	&	 J1235+4109	 	&	 	15.47	 	&	 	1.59	 	&	 	0.041	 	&	 	0.193	 	&	 	$	0.563	\pm	0.053	$ 	&	 $	4333	\pm	136	$ 	&	 $	-4.272	\pm	0.008	$ 	&	 $	10.66	\pm	1.21	$	\\	 
  1570514066627694336	&	 J1250+5446	 	&	 	18.01	 	&	 	1.75	 	&	 	0.004	 	&	 	0.024	 	&	 	$	0.563	\pm	0.038	$ 	&	 $	4033	\pm	97	$ 	&	 $	-4.402	\pm	0.014	$ 	&	 $	11.93	\pm	1.25	$	\\	 
  1531097433767946240	&	 J1255+4655	 	&	 	15.57	 	&	 	1.17	 	&	 	0.011	 	&	 	0.055	 	&	 	$	0.728	\pm	0.112	$ 	&	 $	4641	\pm	286	$ 	&	 $	-4.322	\pm	0.010	$ 	&	 $	9.52	\pm	0.99	$	\\	 
  1459546263999675264	&	 J1303+2603	 	&	 	15.51	 	&	 	1.45	 	&	 	0.011	 	&	 	0.058	 	&	 	$	0.562	\pm	0.041	$ 	&	 $	4293	\pm	106	$ 	&	 $	-4.290	\pm	0.009	$ 	&	 $	11.06	\pm	1.56	$	\\	 
  6085402414245451520	&	 J1312-4728	 	&	 	15.58	 	&	 	1.53	 	&	 	0.002	 	&	 	0.014	 	&	 	$	0.542	\pm	0.002	$ 	&	 $	4126	\pm	19	$ 	&	 $	-4.337	\pm	0.006	$ 	&	 $	12.41	\pm	0.22	$	\\	 
  3607725941130742528	&	 J1316-1536	 	&	 	11.09	 	&	 	-0.20	 	&	 	0.002	 	&	 	0.004	 	&	 	$	0.551	\pm	0.015	$ 	&	 $	15778	\pm	71	$ 	&	 $	-1.991	\pm	0.015	$ 	&	 $	2.37	\pm	0.60	$	\\	 
  6188655210447329792	&	 J1338-2747		&	 	15.68	 	&	 	1.50	 	&	 	0.085	 	&	 	0.302	 	&	 	$	0.594	\pm	0.064	$ 	&	 $	4190	\pm	162	$ 	&	 $	-4.367	\pm	0.021	$ 	&	 $	11.09	\pm	1.46	$	\\	 
  6165095738576250624	&	 J1342-3415	 	&	 	14.66	 	&	 	1.05	 	&	 	0.002	 	&	 	0.009	 	&	 	$	0.521	\pm	0.013	$ 	&	 $	5055	\pm	39	$ 	&	 $	-3.956	\pm	0.010	$ 	&	 $	9.45	\pm	1.86	$	\\	 
  3714266139665215488	&	 J1348+0527	 	&	 	15.03	 	&	 	1.20	 	&	 	0.086	 	&	 	0.233	 	&	 	$	0.566	\pm	0.058	$ 	&	 $	4790	\pm	153	$ 	&	 $	-4.099	\pm	0.022	$ 	&	 $	8.89	\pm	2.25	$	\\	 

\hline \hline
\end{tabular}
\end{center}
\end{table*}

\begin{table*}
\contcaption{ }
\begin{center}
\begin{tabular}{cccccccccc}  
\hline \hline 
Gaia & Short	 & $M_{\rm G}$ &  $G_{\rm{\tiny BP}}\!-\!G_{\rm{\tiny RP}}$ & $\sigma_{M_{\rm G}}$ &  $\sigma_{G_{\rm{\tiny BP}}\!-\!G_{\rm{\tiny RP}}}$ &  Mass & $T_{\rm eff}$ &	Luminosity & Age  \\
Source ID & name     & (mag) & (mag) & (mag) & (mag) & $M_{\odot}$ & (K)  & $\log(L/L_{\odot})$ & (Gyr) \\  
\hline 

  1174809276422844160	&	 J1442+1003	 	&	 	15.51	 	&	 	1.40	 	&	 	0.105	 	&	 	0.324	 	&	 	$	0.593	\pm	0.071	$ 	&	 $	4366	\pm	183	$ 	&	 $	-4.292	\pm	0.026	$ 	&	 $	10.43	\pm	1.58	$	\\	 
  1161215296909017728	&	 J1450+0733	 	&	 	11.14	 	&	 	-0.17	 	&	 	0.004	 	&	 	0.006	 	&	 	$	0.537	\pm	0.015	$ 	&	 $	14638	\pm	180	$ 	&	 $	-2.069	\pm	0.015	$ 	&	 $	3.46	\pm	1.36	$	\\	 
  1294793345366747776	&	 J1500+3600	 	&	 	14.65	 	&	 	1.17	 	&	 	0.034	 	&	 	0.102	 	&	 	$	0.530	\pm	0.031	$ 	&	 $	5077	\pm	89	$ 	&	 $	-3.956	\pm	0.013	$ 	&	 $	9.19	\pm	2.50	$	\\	 
  1600259390916467072	&	 J1502+5409	 	&	 	15.33	 	&	 	1.42	 	&	 	0.060	 	&	 	0.204	 	&	 	$	0.559	\pm	0.044	$ 	&	 $	4466	\pm	113	$ 	&	 $	-4.217	\pm	0.016	$ 	&	 $	10.33	\pm	1.82	$	\\	 
  1612339420228653440	&	 J1503+5509	 	&	 	15.76	 	&	 	1.36	 	&	 	0.073	 	&	 	0.265	 	&	 	$	0.615	\pm	0.083	$ 	&	 $	4163	\pm	210	$ 	&	 $	-4.399	\pm	0.020	$ 	&	 $	11.23	\pm	1.38	$	\\	 
  5824436284328653312	&	 J1517-6645	 	&	 	15.99	 	&	 	1.30	 	&	 	0.102	 	&	 	0.359	 	&	 	$	0.661	\pm	0.114	$ 	&	 $	4029	\pm	307	$ 	&	 $	-4.506	\pm	0.029	$ 	&	 $	11.70	\pm	1.56	$	\\	 
  6007140379167609984	&	 J1518-3803	 	&	 	15.66	 	&	 	1.55	 	&	 	0.040	 	&	 	0.173	 	&	 	$	0.571	\pm	0.043	$ 	&	 $	4151	\pm	108	$ 	&	 $	-4.359	\pm	0.011	$ 	&	 $	11.39	\pm	1.36	$	\\	 
  1277219369981634432	&	 J1522+3146	 	&	 	15.70	 	&	 	1.54	 	&	 	0.103	 	&	 	0.351	 	&	 	$	0.596	\pm	0.066	$ 	&	 $	4171	\pm	170	$ 	&	 $	-4.376	\pm	0.025	$ 	&	 $	11.18	\pm	1.45	$	\\	 
  1277232907719022464	&	 J1523+3152	 	&	 	15.59	 	&	 	1.26	 	&	 	0.021	 	&	 	0.098	 	&	 	$	0.626	\pm	0.079	$ 	&	 $	4364	\pm	197	$ 	&	 $	-4.326	\pm	0.010	$ 	&	 $	10.30	\pm	1.35	$	\\	 
  5827557213731539328	&	 J1539-6124	 	&	 	15.18	 	&	 	1.31	 	&	 	0.026	 	&	 	0.104	 	&	 	$	0.550	\pm	0.039	$ 	&	 $	4593	\pm	100	$ 	&	 $	-4.157	\pm	0.012	$ 	&	 $	10.05	\pm	1.97	$	\\	 
  5817295536128445568	&	 J1707-6319	 	&	 	15.10	 	&	 	1.40	 	&	 	0.019	 	&	 	0.079	 	&	 	$	0.533	\pm	0.029	$ 	&	 $	4637	\pm	74	$ 	&	 $	-4.123	\pm	0.012	$ 	&	 $	10.68	\pm	2.00	$	\\	 
  5802598780807649920	&	 J1715-7323	 	&	 	15.43	 	&	 	1.09	 	&	 	0.111	 	&	 	0.310	 	&	 	$	0.640	\pm	0.116	$ 	&	 $	4586	\pm	304	$ 	&	 $	-4.255	\pm	0.025	$ 	&	 $	9.50	\pm	1.63	$	\\	 
  1711005951573009792	&	 J1749+8247	 	&	 	13.15	 	&	 	0.47	 	&	 	0.001	 	&	 	0.003	 	&	 	$	0.547	\pm	0.000	$ 	&	 $	7257	\pm	0	$ 	&	 $	-3.331	\pm	0.000	$ 	&	 $	4.33	\pm	0.00	$	\\	 
  6363668569344689408	&	 J1812-8028	 	&	 	15.55	 	&	 	1.04	 	&	 	0.143	 	&	 	0.390	 	&	 	$	0.659	\pm	0.131	$ 	&	 $	4520	\pm	353	$ 	&	 $	-4.302	\pm	0.029	$ 	&	 $	9.85	\pm	1.54	$	\\	 
  6653858618815379328	&	 J1814-5305	 	&	 	15.68	 	&	 	1.08	 	&	 	0.130	 	&	 	0.376	 	&	 	$	0.669	\pm	0.132	$ 	&	 $	4396	\pm	356	$ 	&	 $	-4.361	\pm	0.027	$ 	&	 $	10.41	\pm	1.47	$	\\	 
  4484289866726156160	&	 J1824+1213	 	&	 	15.57	 	&	 	1.63	 	&	 	0.012	 	&	 	0.069	 	&	 	$	0.548	\pm	0.025	$ 	&	 $	4195	\pm	74	$ 	&	 $	-4.316	\pm	0.011	$ 	&	 $	11.72	\pm	1.25	$	\\	 
  2146619161278293248	&	 J1852+5333	 	&	 	15.72	 	&	 	1.39	 	&	 	0.134	 	&	 	0.410	 	&	 	$	0.617	\pm	0.093	$ 	&	 $	4216	\pm	245	$ 	&	 $	-4.379	\pm	0.035	$ 	&	 $	11.06	\pm	1.44	$	\\	 
  6663268308043562112	&	 J1926-4627	 	&	 	15.02	 	&	 	1.46	 	&	 	0.060	 	&	 	0.190	 	&	 	$	0.533	\pm	0.029	$ 	&	 $	4711	\pm	79	$ 	&	 $	-4.095	\pm	0.016	$ 	&	 $	10.25	\pm	2.06	$	\\	 
  6647162730439433984	&	 J1936-4913	 	&	 	15.58	 	&	 	1.50	 	&	 	0.057	 	&	 	0.219	 	&	 	$	0.574	\pm	0.047	$ 	&	 $	4251	\pm	121	$ 	&	 $	-4.320	\pm	0.015	$ 	&	 $	10.94	\pm	1.47	$	\\	 
  2301882675705225472	&	 J1940+8348	 	&	 	15.51	 	&	 	1.23	 	&	 	0.010	 	&	 	0.049	 	&	 	$	0.608	\pm	0.065	$ 	&	 $	4413	\pm	155	$ 	&	 $	-4.290	\pm	0.012	$ 	&	 $	9.78	\pm	1.39	$	\\	 
  2082254987541672960	&	 J2006+4544	 	&	 	15.21	 	&	 	1.22	 	&	 	0.037	 	&	 	0.136	 	&	 	$	0.574	\pm	0.057	$ 	&	 $	4629	\pm	143	$ 	&	 $	-4.169	\pm	0.013	$ 	&	 $	9.32	\pm	1.93	$	\\	 
  6471523921227261056	&	 J2042-5218	 	&	 	14.68	 	&	 	0.94	 	&	 	0.043	 	&	 	0.121	 	&	 	$	0.601	\pm	0.070	$ 	&	 $	5249	\pm	199	$ 	&	 $	-3.973	\pm	0.015	$ 	&	 $	6.22	\pm	1.96	$	\\	 
  1737588947276271744	&	 J2052+0709	 	&	 	15.89	 	&	 	1.56	 	&	 	0.059	 	&	 	0.250	 	&	 	$	0.605	\pm	0.061	$ 	&	 $	3980	\pm	158	$ 	&	 $	-4.468	\pm	0.020	$ 	&	 $	11.78	\pm	1.38	$	\\	 
  6580458035746362496	&	 J2117-4156a	 	&	 	15.15	 	&	 	1.33	 	&	 	0.024	 	&	 	0.095	 	&	 	$	0.542	\pm	0.033	$ 	&	 $	4608	\pm	84	$ 	&	 $	-4.143	\pm	0.012	$ 	&	 $	10.28	\pm	1.96	$	\\	 
  6580551872194787968	&	 J2117-4156b	 	&	 	15.34	 	&	 	1.30	 	&	 	0.028	 	&	 	0.115	 	&	 	$	0.572	\pm	0.052	$ 	&	 $	4493	\pm	130	$ 	&	 $	-4.220	\pm	0.013	$ 	&	 $	9.94	\pm	1.78	$	\\	 
  1783614400935169408	&	 J2127+1545	 	&	 	14.30	 	&	 	0.82	 	&	 	0.023	 	&	 	0.066	 	&	 	$	0.599	\pm	0.058	$ 	&	 $	5700	\pm	191	$ 	&	 $	-3.827	\pm	0.012	$ 	&	 $	5.06	\pm	1.86	$	\\	 
  2687584757658775424	&	 J2129-0034	 	&	 	15.37	 	&	 	1.57	 	&	 	0.112	 	&	 	0.341	 	&	 	$	0.562	\pm	0.045	$ 	&	 $	4427	\pm	120	$ 	&	 $	-4.236	\pm	0.022	$ 	&	 $	10.46	\pm	1.76	$	\\	 
  6465689878168451328	&	 J2139-5058	 	&	 	15.85	 	&	 	1.55	 	&	 	0.084	 	&	 	0.320	 	&	 	$	0.607	\pm	0.067	$ 	&	 $	4025	\pm	175	$ 	&	 $	-4.450	\pm	0.025	$ 	&	 $	11.68	\pm	1.33	$	\\	 
  2205493129867600256	&	 J2225+6357	 	&	 	14.53	 	&	 	1.05	 	&	 	0.006	 	&	 	0.024	 	&	 	$	0.525	\pm	0.020	$ 	&	 $	5190	\pm	63	$ 	&	 $	-3.912	\pm	0.012	$ 	&	 $	8.98	\pm	2.20	$	\\	 
  6357629089412187648	&	 J2230-7515	 	&	 	15.57	 	&	 	1.55	 	&	 	0.002	 	&	 	0.012	 	&	 	$	0.575	\pm	0.059	$ 	&	 $	4275	\pm	157	$ 	&	 $	-4.313	\pm	0.011	$ 	&	 $	10.61	\pm	0.99	$	\\	 
  2709539840202060800	&	 J2237+0636	 	&	 	15.85	 	&	 	1.54	 	&	 	0.110	 	&	 	0.396	 	&	 	$	0.617	\pm	0.081	$ 	&	 $	4058	\pm	211	$ 	&	 $	-4.446	\pm	0.031	$ 	&	 $	11.60	\pm	1.37	$	\\	 
  1941133391670459648	&	 J2314+4545	 	&	 	15.96	 	&	 	1.41	 	&	 	0.099	 	&	 	0.358	 	&	 	$	0.640	\pm	0.095	$ 	&	 $	4008	\pm	248	$ 	&	 $	-4.492	\pm	0.029	$ 	&	 $	11.74	\pm	1.44	$	\\	 
  2631967439437024384	&	 J2319-0613	 	&	 	15.06	 	&	 	1.37	 	&	 	0.007	 	&	 	0.035	 	&	 	$	0.530	\pm	0.023	$ 	&	 $	4661	\pm	52	$ 	&	 $	-4.112	\pm	0.013	$ 	&	 $	10.40	\pm	1.68	$	\\	 
  2641576685735609472	&	 J2349-0124	 	&	 	14.99	 	&	 	1.22	 	&	 	0.028	 	&	 	0.099	 	&	 	$	0.544	\pm	0.038	$ 	&	 $	4771	\pm	97	$ 	&	 $	-4.083	\pm	0.013	$ 	&	 $	9.43	\pm	2.17	$	\\	 
  2310942857676734848	&	 J2354-3634	 	&	 	11.29	 	&	 	-0.16	 	&	 	0.003	 	&	 	0.005	 	&	 	$	0.574	\pm	0.018	$ 	&	 $	14636	\pm	121	$ 	&	 $	-2.134	\pm	0.005	$ 	&	 $	1.68	\pm	0.49	$	\\				 
\hline \hline
\end{tabular}
\end{center}
\end{table*}

\subsection{Completeness analysis of the halo white dwarf sample}
\label{ss:comple}

\begin{figure}
      {\includegraphics[width=1.0\columnwidth, clip=true,trim=15 2 15 30]{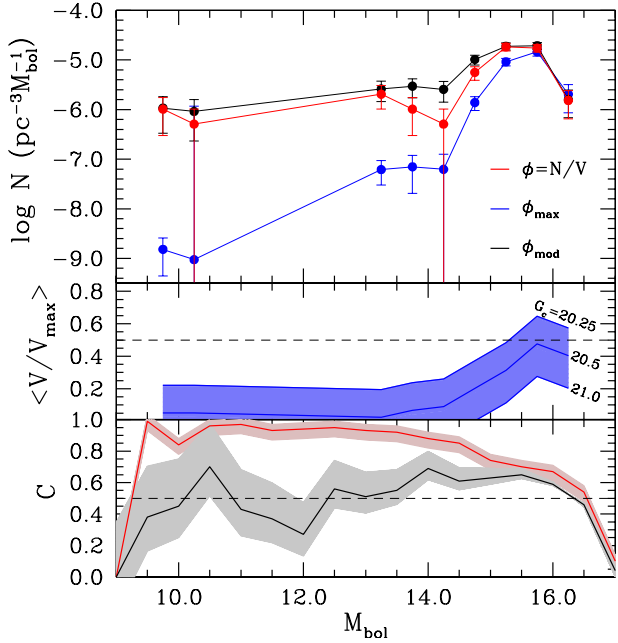}}
   \caption{Top panel: Halo white dwarf luminosity functions built with the classical estimator $\phi=N/V$ (red line and dots), with the $1/\mathcal{V}_{\rm max}$ method, $\phi_{\rm max}$ (blue line and dots), and with the modified volume, $\phi_{\rm mod}$ (black line and dots). Middle panel: average $\langle V/V_{\rm max}\rangle$ as a function of the bolometric magnitude for different magnitude cuts, $G_c\in[20.25,21.0]\,$mag. Also shown as a dashed line the value representing an uniform complete sample, $\langle V/V_{\rm max}\rangle=0.5$. Bottom panel: completeness of the whole {\it Gaia} 100\,pc sample of white dwarfs (red line, $1\sigma$ pink area) and that of the halo subsample (black line, $1\sigma$ gray area).} 
  \label{f:comple}
\end{figure}

As a preliminary step for building the halo white dwarf luminosity function, we analyze in detail the completeness of our selected sample. First, we need to take into account that our ‘halo sample’ is not a ‘direct’ observed sample. That is, we first obtain a sample of white dwarfs within 100 pc, and from that we apply a classifying method based on a Random Forest algorithm in order to obtain the ‘halo sample’. Then, the incompleteness raises from two sources: from the selection process of the observed 100 pc sample  and from the classification algorithm.\\

With respect to the first item, the {\it Gaia} mission is expected to be (understood as an end-of-mission goal) complete up to $G=21\,$mag. Recent estimates based on the EDR3-{\it Gaia} catalogue of nearby stars  provide a completeness of 97\%, 95\% and 91\% for $G$ magnitudes 19.9, 20.2 and 20.5, respectively \citep{Gaiacol2020}. Taking this into account, we extend our completeness analysis of the 100\,pc sample (see Figure  5 from \cite{Jimenez2018}) up to $G_{BP}-G_{RP}=2.0$. The resulting completeness as a function of the bolometric magnitude for the whole {\it Gaia} 100\,pc sample is shown in the bottom panel (red line) of Figure \ref{f:comple}. A completeness above 90\% is achieved for most of the sample up to $M_{\rm bol}\sim14.0$. At this magnitude, we note a decreasing trend reaching a 50\% completeness at magnitude $M_{\rm bol}\sim16.6\,$mag. 

Secondly, we analyze the incompleteness introduced by our  Random Forest classification algorithm. We recall that in our classification process we used an 8-dimensional space  (equatorial coordinates, parallax, proper motion components and photometric magnitudes), where the algorithm estimates the  entropy function, evaluated on each splitting branch of the different decisions trees of the Random Forest algorithm (see \citealt{Torres2019}).  For each individual object the minimum entropy found, $S_i$, is used to classify the object in a certain Galactic component. Consequently, this entropy value $S_i$ can be logically understood as an inverse of the probability that an object belongs to a certain group. Using our Monte Carlo simulator we generate several samples of $\sim95$ halo white dwarfs and calculate, through our Random Forest algorithm, the individual probability of being classified as such. We compute, then, the average value as a function of the bolometric magnitude taking into account the  completeness of the whole 100\,pc sample. The resulting distribution is shown in the bottom panel of Figure \ref{f:comple} as a black line (gray area corresponding to $1\sigma$ dispersion). We observe an irregular pattern for the bright region ($M_{\rm bol}\lppr 14\,$mag), while for fainter magnitudes the halo white dwarf completeness distribution resembles that of the whole sample. According to our analysis, the noisy appearance of the bright region is due to two factors: mainly, the low number  of objects expected in these bins, which are dominated by Poisson counting error statistics, and secondly, and  to a minor extent, to the intrinsic difficulty to classify a bright object as belonging to the halo population. High-speed and cool objects are closer to what we expect to be a halo white dwarf than hot and high-speed objects. In the first case, the automatic classification algorithm seems to relax the kinematic condition to accept moderately fast objects if they are cool enough, while the reverse condition happens for hot halo candidates. Besides, more hot high-speed disk objects may contaminate the hot region and the difficulty to disentangle halo objects increases in that region.  On the other hand, for fainter magnitudes the halo sample completeness resembles that of the whole 100\,pc sample, indicating that the probability to correctly classifying a halo object in this region is very high. It should be noted that the completeness of the halo sample for magnitudes around $M_{\rm bol}=16.0$ is $\sim60\%$, which can be considered as an acceptable value, while for the rest of the sample is around $\sim50\%$. 

We are in position now to estimate the luminosity function. Among the different estimators proposed (see, for instance, \citealt{Geijo2006} and reference therein) the $1/\mathcal{V}_{\rm max}$ method \citep{Schmidt1968,Felten1976} is the most commonly used to estimate the white dwarf luminosity function. However, this method,  based on the assumption of a complete magnitude-limited sample, is not applicable in our case, since our classification method introduces an incompleteness factor which is not only magnitude dependent. Consequently, we will use the $1/\mathcal{V}_{\rm max}$ method for comparative purposes just as a first guess in our analysis. In the middle panel of Figure \ref{f:comple} we show the average $\langle V/V_{\rm max}\rangle$ as function of $M_{\rm bol}$ adopting different $G$ magnitude cuts, $G_{c}$, of the whole 100\,pc sample in the range $G_c\in[20.25,21.0]\,$mags. We also recall that the $1/\mathcal{V}_{\rm max}$ provides an estimation of the completeness of the sample, once assumed a complete uniformly distributed sample, resulting in a average value of $\langle V/V_{\rm max}\rangle=0.5$. In our case, adopting a conservative value of $G_c=20.5\,$mags, the $\langle V/V_{\rm max}\rangle$ distribution is closer to the value 0.5 (dashed line) for magnitudes $M_{\rm bol}\sim15.7\,$mags. Specifically, a value of  $\langle V/V_{\rm max}\rangle=0.476\pm0.049$ (adopting for $N$ objects a deviation of $1/\sqrt{12N}$, \citealt{Rowell2011}) thus indicating an acceptable degree of completeness for that particular magnitude bin. However, as we mentioned earlier, brighter magnitudes appear to be far from being complete as shown by the low $\langle V/V_{\rm max}\rangle$ value.

In the top panel of Figure \ref{f:comple}, we compare the different halo white dwarfs luminosity functions built from our previous completeness analysis. First, we use the classical estimator of the luminosity function for volume-limited samples, $\phi=N/V$. The resulting luminosity function is shown in red. Second, the $1/\mathcal{V}_{\rm max}$ provides an estimation of the luminosity function, $\phi_{\rm max}$,  binning the sample in $i\in(1, N)$ magnitude bins and weighting the contribution of each object as inversely proportional to its maximum volume available within the selection cuts: $\phi_{\rm max}=\sum_{j=1}^{N_i}1/\mathcal{V}^j_{\rm max}$, where the deviation is calculated as the sum in quadrature of the individual errors. Adopting a magnitude cut of $G_c=20.5\,$mag, the luminosity function thus obtained is shown in blue in the top panel of Fig. \ref{f:comple}. However, as previously stated, both, classical and  $1/\mathcal{V}_{\rm max}$ method, can be only understood as a first guess for building a proper luminosity function. 

In order to properly take into account all possible sources of incompleteness in our sample, as previously discussed, we use a generalization of the  $1/\mathcal{V}_{\rm max}$ method (\citealt{Lam2015}  and references therein). A modified volume is defined as
\begin{equation}
    \mathcal{V}^i_{{\rm mod}}=\Omega_i\int^{r_{\rm max}}_{r_{\rm min}}\chi\frac{\rho(r)}{\rho_{\sun}} r^2dr
\end{equation}
where $\Omega_i$ is the solid angle covered by object $i$, $r_{\rm max}$ and $r_{\rm min}$ are, respectively the maximum and minimum distances and $\rho(r)/\rho_{\sun}$  the density ratio along the line of side. This volume integral is modified by the function $\chi$, traditionally named as the discovery fraction, and typically used for taking into account the number of objects that pass a certain tangential velocity threshold. In our case, we generalize the $\chi$ function to incorporate our 8-dimensional space of equatorial coordinates, proper motion components, parallax and photometric magnitudes, $\chi=\chi(\alpha,\delta,\mu^*_{\alpha},\mu_{\delta},\varpi, G,G_{RP},G_{BP})$. Thus, the $\chi$ function becomes equivalent to the probability derived from our Random Forest algorithm of being correctly classified as an halo member. Adopting a constant density profile, which is a natural assumption for a halo distribution in a small volume, we can compute the luminosity function as a generalization of the $1/\mathcal{V}_{\rm max}$ method by using the previously modified volume: $\phi_{\rm mod}=\sum_{j=1}^{N_i}1/\mathcal{V}^j_{\rm mod}$. The corresponding luminosity function, $\phi_{\rm mod}$ is shown in black in the top panel of Fig. \ref{f:comple}.

The analysis of the three estimators studied here reveals a similar trend of the corresponding luminosity functions for the faint region, i.e. $M_{\rm bol}\gappr 15\,$mag, being this fact a consequence of a reasonable completeness ($\sim60\%$) for this low-luminosity region. On the contrary, the brighter region suffers from a larger degree of incompleteness, mainly due to a low number statistic. In this situation,  the $1/\mathcal{V}_{\rm max}$ is markedly inefficient at recovering the original distribution,  greatly underestimating the brightest bins. Given that our modified estimator is the only one that properly takes into account the different sources of incompleteness, in what follows we will use the luminosity function derived from it as representative of the halo white dwarf population.

\begin{figure*}
      {\includegraphics[width=1.0\columnwidth, clip=true,trim=5 20 15 30]{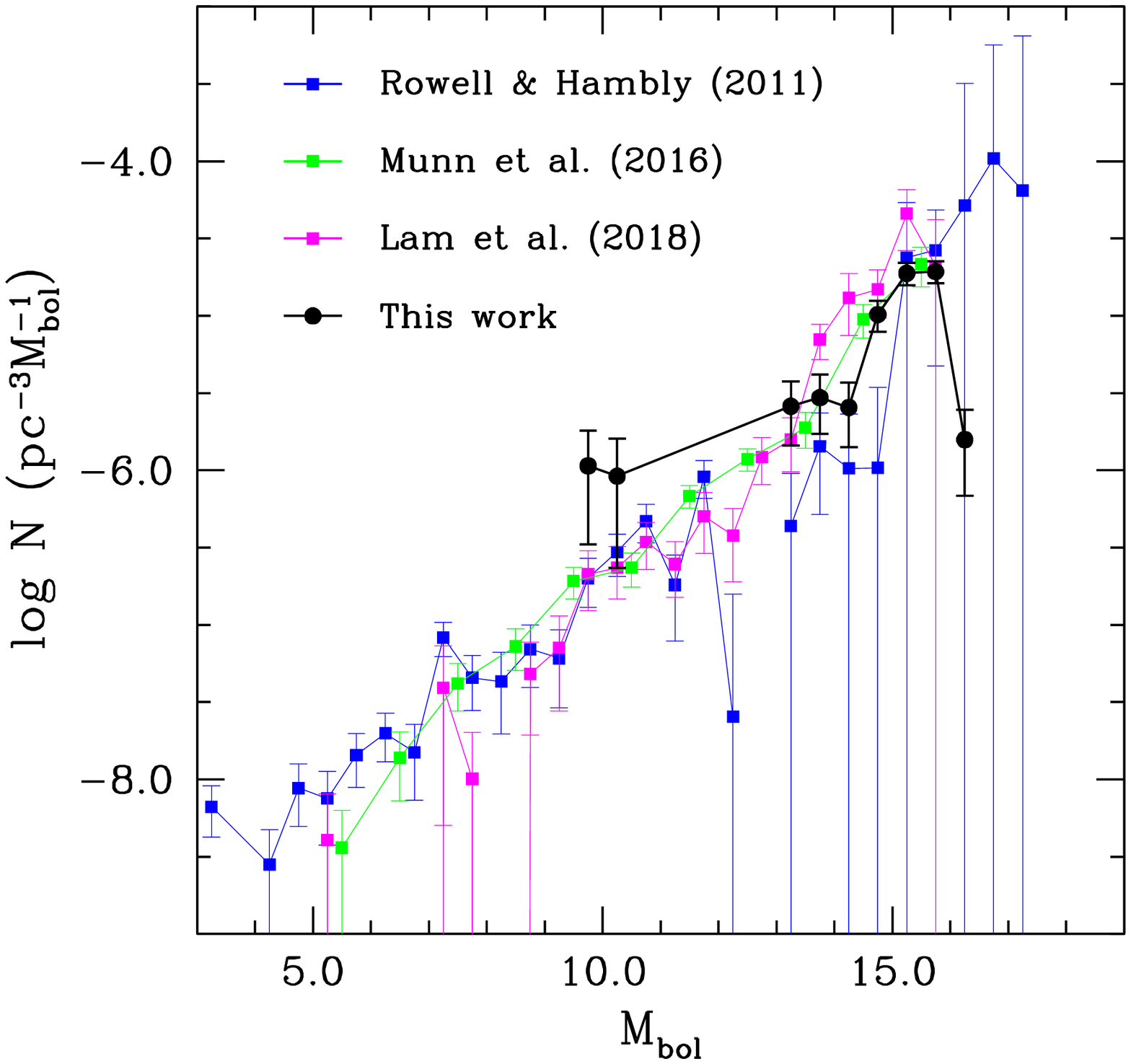}}
       {\includegraphics[width=1.0\columnwidth, clip=true,trim=5 20 15 30]{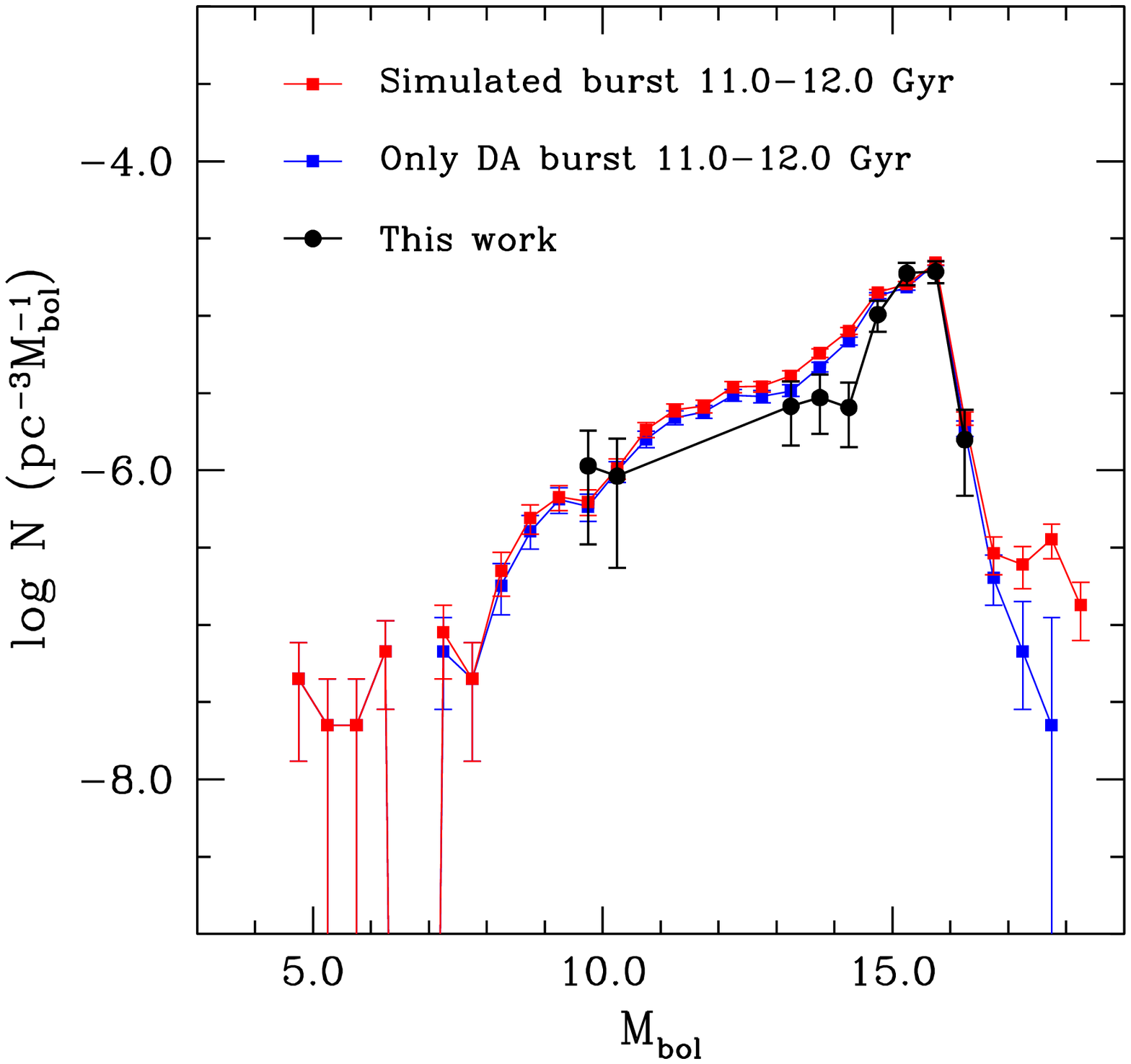}}
  \caption{Left panel: {\it Gaia} halo white dwarf luminosity function within 100\,pc obtained in this work (black dots and lines) compared to the \citet{Rowell2011} spheroid white  dwarf luminosity function (blue squares and lines), and the luminosity function for high speed white dwarfs obtained by \citet{Munn2016} (green squares and lines) and \citet{Lam2018} (magenta squares and lines). Right panel: {\it Gaia} halo white dwarf luminosity function obtained in this work (black dots and lines), compared with the synthetic populations simulated in this work that best fit the observed features. This is, synthetic populations considering a star formation burst from 11 to 12 Gyrs ago, and a standard proportion of DA to DB white dwarfs of 80-20\% (red squares and lines) and one in which all synthetic white dwarfs are DA (blue squares and lines).}
  \label{f:lf}
\end{figure*}

\subsection{The halo white dwarf luminosity function}
\label{ss:hwdlf}

In Figure \ref{f:lf} (left panel) we plot the luminosity function derived in this work (black) compared to some of the most recent white dwarf luminosity functions obtained for a high speed or equivalently halo population. In particular, we show the spheroid white dwarf luminosity function obtained from the SuperCOSMOS Sky Survey via the effective volume technique from \cite{Rowell2011} --blue squares--; the luminosity function for high speed white dwarfs, $200<v_{\rm tan}<500\,$\kms, derived from SDSS  deep proper motion survey by \cite{Munn2016} -- green squares --; and the most recent high speed white dwarf luminosity function derived by \cite{Lam2018} from the Pan–STARRS 1 3$\pi$ Steradian Survey -- magenta squares. 

First of all, the space density previously derived in \citet{Torres2019} for our halo candidate sample, $(4.8\pm0.4)\times10^{-5}\,{\rm pc^{-3}}$,   is in agreement with the value reported by \cite{Lam2018}, $(5.291\pm0.2)\times10^{-5}\,{\rm pc^{-3}}$. Moreover, our space density estimate is slightly larger than that of \cite{Munn2016}, $(3.5\pm0.7)\times10^{-5}\,{\rm pc^{-3}}$, but it is well below the upper limit of $(1.9\times10^{-4}\,{\rm pc^{-3}}$ presented by \cite{Rowell2011}. 

Second, our white dwarf luminosity function is defined in fewer magnitude bins than the rest of samples. That paucity in the number of objects is a consequence of the relative small size of our sample, which is limited within 100\,pc. However, we need to recall that our sample has been extracted from  a nearly volume limited -- and thus practically complete -- sample, while the rest of samples are magnitude-limited (see Section \ref{ss:comple} for a detailed analysis of the completeness of our sample). Consequently some biases are expected, in particular for the dimmer intervals of the luminosity function. In this sense, our luminosity function shows a clear peak at $M_{\rm bol}\sim15.5$ mag and a marked drop-off for the fainter bins. The location of the peak agrees with the one presented by the \cite{Lam2018}'s luminosity function. At the same time, the slight depression shown at $M_{\rm bol}\sim14.5$ mags by our luminosity function is also present in that of \cite{Rowell2011} but seems not to appear in any of the other two distributions. However, the drop-off beyond 15.5 mags is the most relevant feature of our luminosity function, which is not present in any other of the distributions. The only white dwarf luminosity function which extends beyond the cut-off  is that of \cite{Rowell2011}. Unfortunately, their faintest bins are poorly constrained and no cut-off is observed.  Based on the completeness analysis of our sample (see Section \ref{ss:comple}), we can claim that this drop-off is real and thus we are observing for the first time the cut-off of the halo white dwarf luminosity function.

Fitting the cut-off of the white dwarf luminosity function has been extensively used as a consistent technique for estimating ages  \citep[e.g][]{GBerroOswalt2016}. For this purpose we build a set of synthetic white dwarf luminosity functions derived from a 1 Gyr burst of star formation applied at different ages. Our best fit, that is, the one that best reproduces the cut-off of our luminosity function, corresponds to a burst that happened between 11.0-12.0 Gyr in the past. In the right panel of Fig. \ref{f:lf} we show our results for the best fit model when a population of 80:20, DA to non-DA ratio, is considered (red) and when only DAs are taken into account (blue). Both models are able to correctly fit the peak and  the cut-off bins. However, the depression previously commented at $M_{\rm bol}\sim14.5$ mags seems not to be reproduced by a burst model. The paucity of objects in that bin prevents us from drawing further conclusion.  Finally, the possible effects of non-DA stars (in particular those with He-rich atmospheres) are only apparent for the faintest bins of the luminosity function, but these are beyond our observed sample.

\subsection{The mass and age distributions and the star formation history}

\begin{figure*}
      {\includegraphics[width=1.0\columnwidth, clip=true,trim=5 20 15 30]{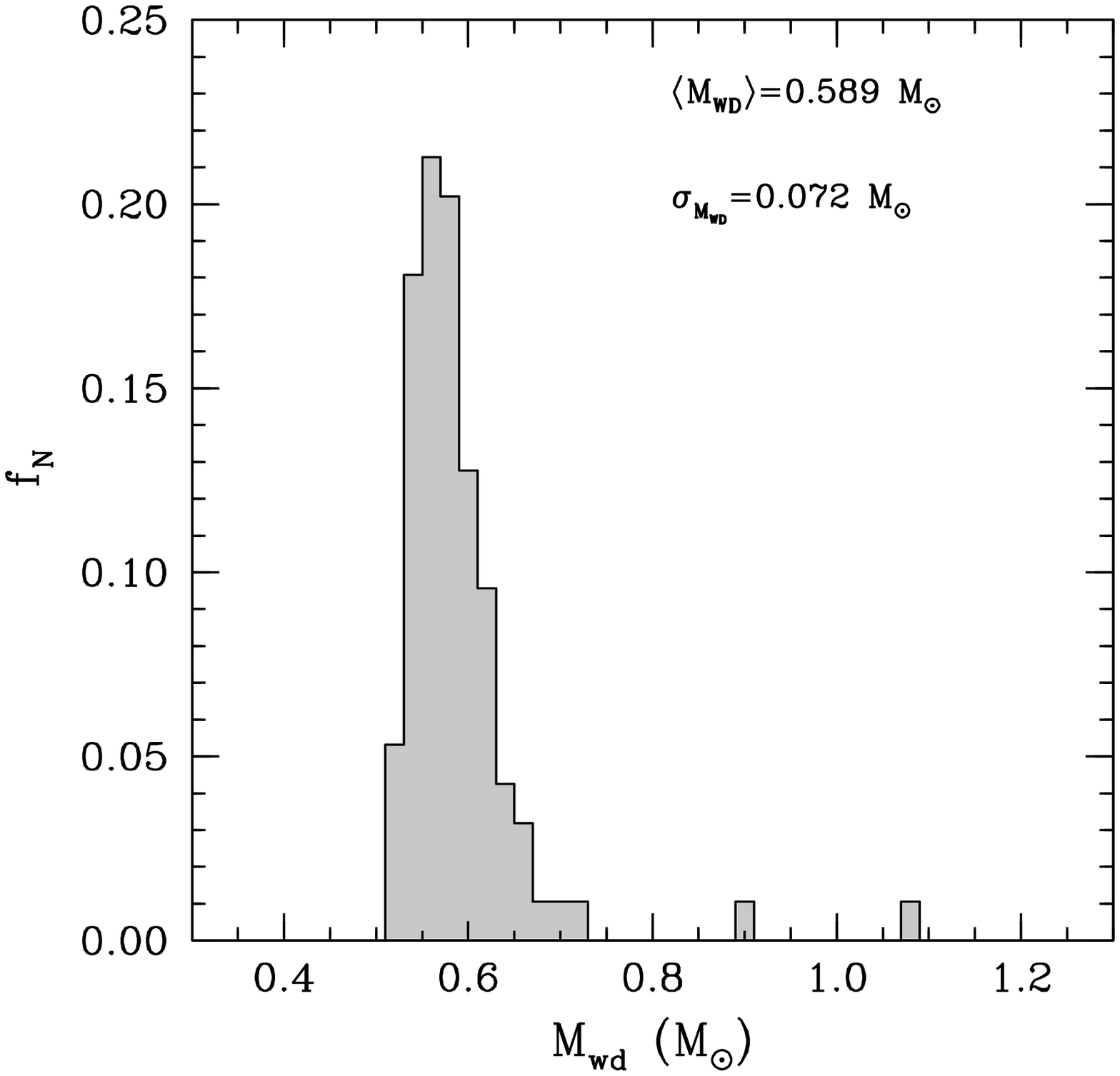}}
       {\includegraphics[width=1.0\columnwidth, clip=true,trim=5 20 15 30]{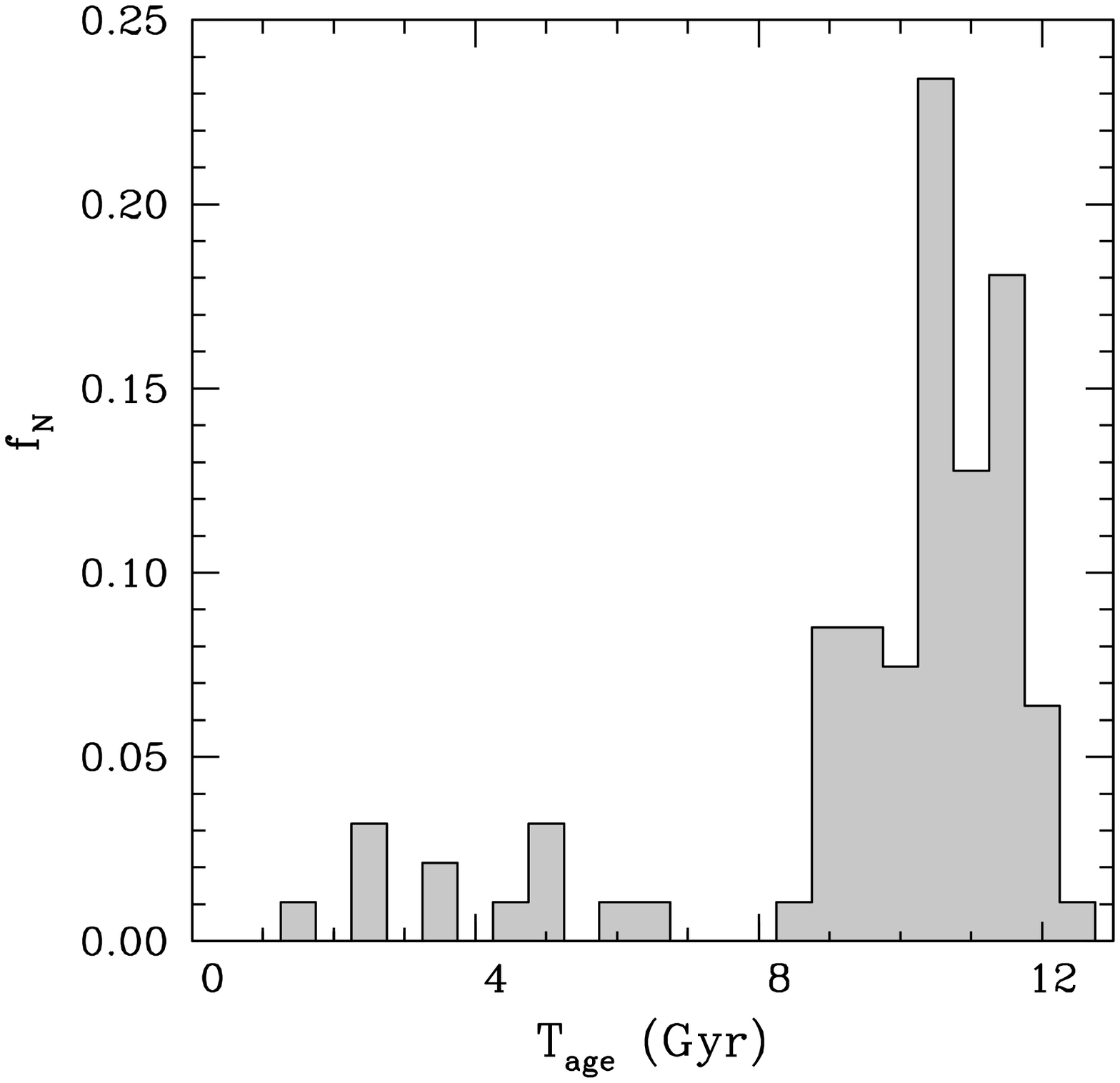}}
  \caption{Left panel: {\it Gaia} halo white dwarf mass function within 100\,pc obtained in this work. Right panel: total age (cooling age plus progenitor lifetime) distribution for our halo white dwarf sample.} 
  \label{f:mt}
\end{figure*}

The photometric method outlined in Section\,\ref{ss:methtest} has also allowed us to derive some other important stellar parameters such as the mass and the age of the white dwarfs. In Figure \ref{f:mt} we show the mass distribution (left panel) for our halo white dwarf sample. The mean mass $\langle M_{\rm WD}\rangle=0.589\,M_{\odot}$ is smaller than the mean value of $\sim0.65\,$ generally reported for the single white dwarf disk population \citep[e.g.][]{Tremblay2016,Bergeron2019,McCleery2020}. The majority of white dwarfs ($71\%$) present a mass smaller than $0.6\,M_{\odot}$, which is expected for an old population, as more low-mass progenitors have had enough time to evolve and become white dwarfs of slightly lower masses than the canonical value. On the other hand, we have found just two massive white dwarfs (J1045+5904 and J1123+5742) with masses $1.074\pm0.033\,M_{\odot}$ and $0.893\pm0.100\,M_{\odot}$, respectively.

In the right panel of Fig. \ref{f:mt} we show the age distribution for our sample of white dwarfs. We recall that the age represented corresponds to the total age of the white dwarfs, that is, the cooling time plus the progenitor lifetime. A first glance to the age distribution reveals that the vast majority of stars ($87\%$) have total ages between 8 to 12\,Gyr. In particular, most of them between 10-12\,Gyr, being compatible with the 11-12\,Gyr  burst we applied for fitting the cut-off of the luminosity function (See section \ref{ss:hwdlf}).  Three objects found in our sample (J0045-3329, J1049-7559 and J1312-4728)  are older than 12\,Gyr, being the  last of them the  oldest with an age of  $12.41\pm0.22\,$Gyr. These stars appear older than some of the previously published oldest white dwarfs in the Solar neighborhood. For instance, the objects  SDSS J1102+4113 and WD 0346+246 have, respectively, age estimates of 11 and 11.5\,Gyr \citep{Kilic2012}. It is worth mentioning that this last object, WD 0346+246 -- which was firstly discover by \cite{Hambly1997} and extensively studied by e.g.  \citet[][]{Oppenheimer2001,Bergeron2001,Kilic2012} -- is also present in our sample with an estimate total age of $11.49\pm1.51\,$Gyr that is in perfect agreement with these previous works.

On the other hand, 12 white dwarfs have age estimates younger than 7\,Gyr. In principle, they may be ruled out as genuine halo members but, at the same time, their high speed kinematics are indicative that their origin is different from the bulk of the disk population. They represent $13\,$\% of the sample, a value that can be considered as a general estimate  of the contamination of any possible halo white dwarf sample. Moreover, these objects seemed to be clustered at around 2.5 and 5\,Gyr, respectively. With the aid of our Monte Carlo simulator, we analyze the statistical significance of this apparent clustering. We uniformly distribute 12 objects in the range 0-8\,Gyr in intervals of 0.5\,Gyr. The probability of obtaining a peak (of three or more stars) is rather high, 0.42. However, the probability of recovering two of these peaks is considerably low, 0.05. In other words, the observed distribution has a statistical significance of $2\sigma$ of rejecting that they come from a uniform distribution. In the forthcoming analysis some hypothesis about their origin are  presented.

\begin{figure}
      {\includegraphics[width=1.0\columnwidth, clip=true,trim=5 20 15 30]{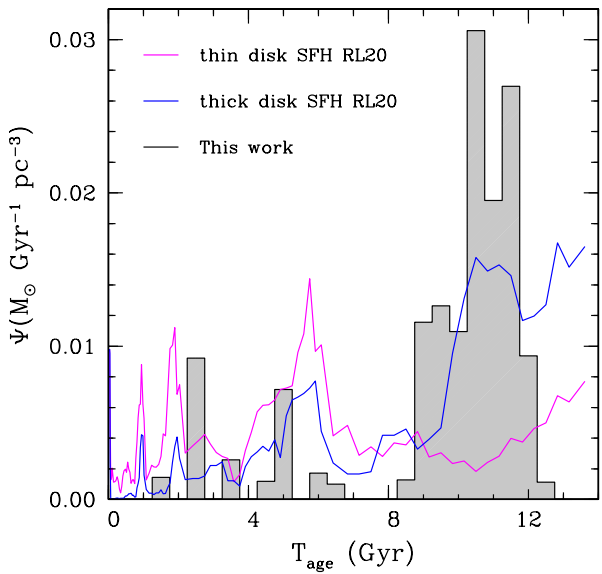}}
  \caption{Star formation rate (gray histogram) obtained in this work for the {\it Gaia} halo white dwarfs within 100\,pc. For comparative purposes we also plot the star formation history for the thin disk (magenta line) and thick disk (blue line) population of the Milky Way determined by \citet{RuizLara2020}. These last two distributions are arbitrarily normalized.} 
  \label{f:sfr}
\end{figure}

\begin{figure*}
      {\includegraphics[width=1.0\columnwidth, clip=true,trim=5 15 5 30]{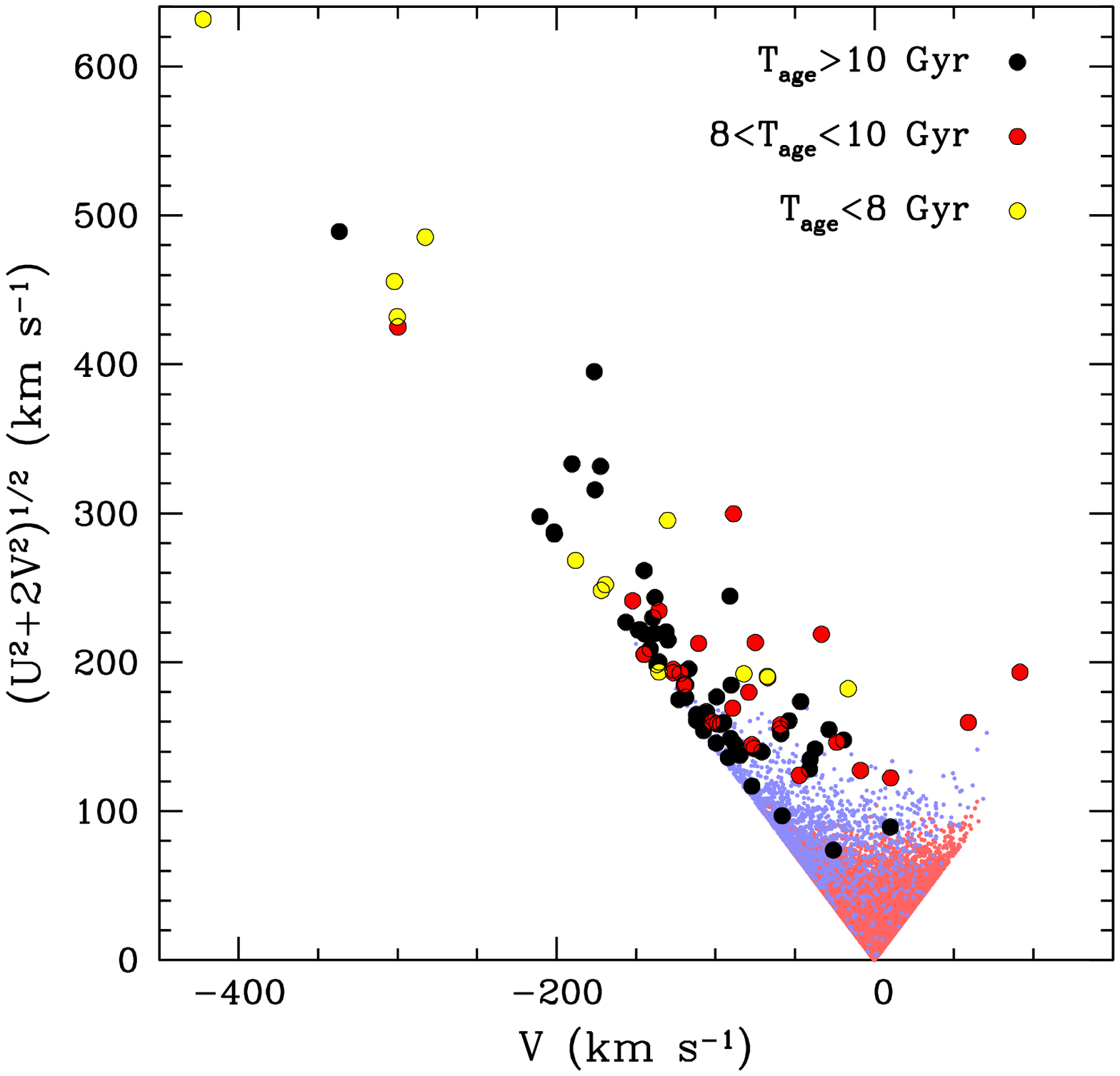}}
        {\includegraphics[width=1.0\columnwidth,clip=true,trim=5 15 5 30]{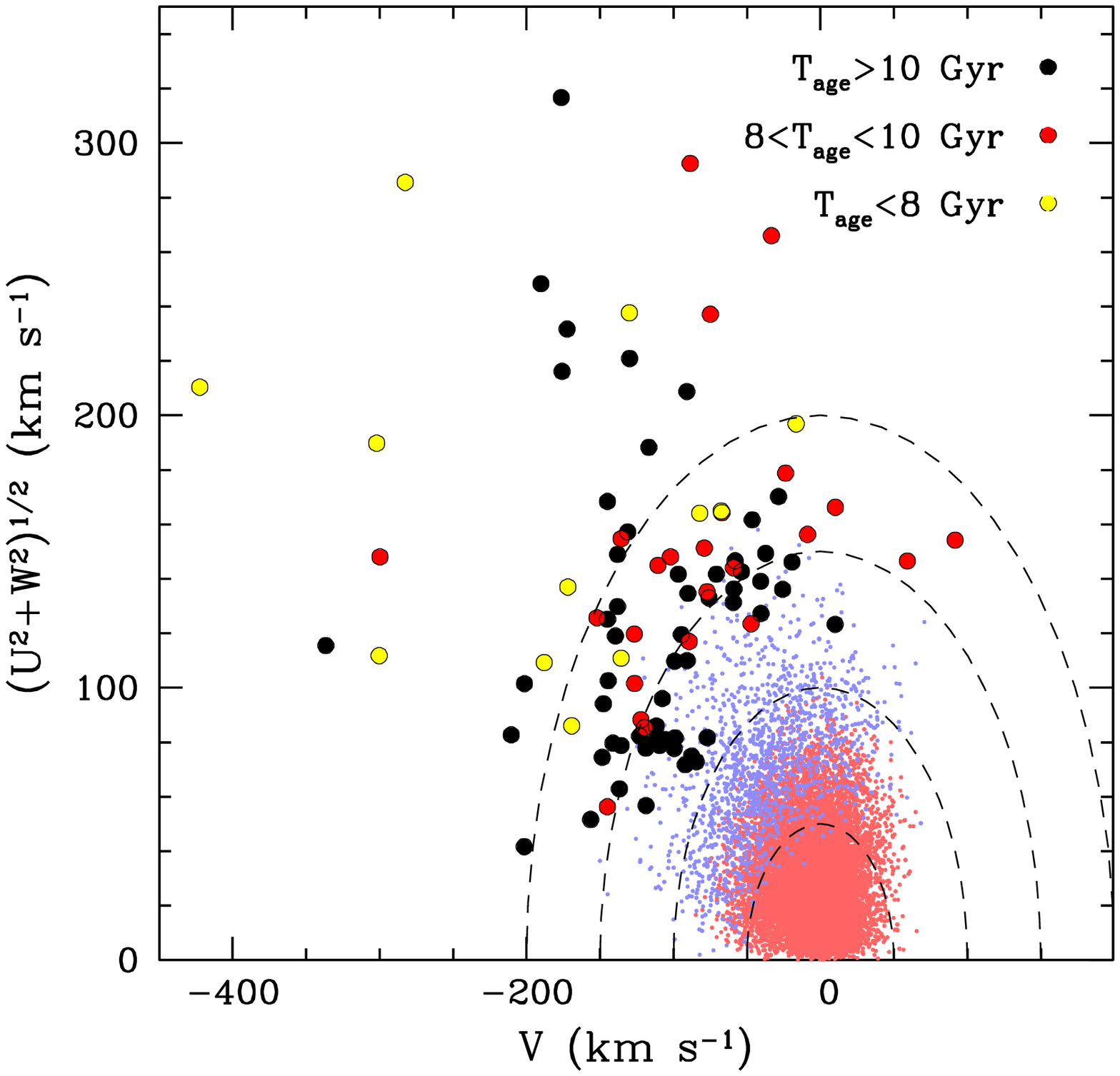}}  
  \caption{Action space corresponding to the integral of motion $V_{\Delta E}\equiv(U^2+2V^2)^{1/2}$ as a function of $V$ (left panel) and the Toomre diagram (right panel) for the white dwarfs of our halo candidate sample (coloured scaled as a function of the age). Also shown, for illustrative purposes, are the thin (light red dots) and thick (light blue dots) velocities for the 100\,pc white dwarf sample from \citet{Torres2019}. Dashed lines represent curves for constant $U^2+V^2+W^2$ at values 50, 100, 150 and 200\kms.} 
  \label{f:kin}
\end{figure*}

A subproduct of our photometric procedure is the mass of the white dwarf progenitor. Given that we also know the age distribution, we can easily estimate the star formation rate, $\Psi$, as the quantity of stellar mass produced per unit time and unit volume. In Figure \ref{f:sfr} we show as a gray histogram our SFR thus computed in units of $M_{\odot}$/Gyr/pc$^3$. The SFR resembles, as logically expected, the age distribution previously shown in Fig.  \ref{f:mt}, although the peaks, in particular for the younger bins, are more pronounced. We observe that the peak of the star formation history is centered at around 11\,Gyr, which is compatible with the current age of the Gaia-Enceladus encounter. It is worth noting here that our estimate of the halo SFR is restricted to the mass range of those progenitors that, according to our population synthesis model,  are able to form a white dwarf, i.e., masses in the range $\sim1-5.5\,M_{\odot}$.

For a comparative purpose we also show in Fig. \ref{f:sfr} the recent star formation history from \cite{RuizLara2020} corresponding to the thin disk (magenta line) and thick disk (blue line) of our Galaxy. In their work, \cite{RuizLara2020} thoroughly analyzed the HR-diagram derived from the precise astrometric and photometric data provided by {\it Gaia} DR2. They found clear evidences that the close encounter of the Milky Way with Saggitarius dwarf galaxy has enhanced the star formation rate  at epochs  5.7, 1.9 and 1 Gyr in the past. The location of the two older peaks seems to resemble the location of the peaks for the younger white dwarfs found in our sample at ~5 and ~2.5\,Gyr. It can be expected that the close encounter with the Saggitarius dwarf galaxy may have an effect not only in the enhancement of star formation but also in heating the kinematics of the affected stars. Consequently, it is reasonable to link the origin of these high-speed and young white dwarfs to the gravitational effects induced by the pass of the Saggitarius galaxy. 

Following the analysis by \cite{RuizLara2020} we also found an enhancement in the star formation of the thick disk population centered at $\sim10.5\,$Gyr.  This enhancement is not associated to the encounters with the Saggitarius dwarf galaxy, but rather to the formation of the thick disk itself. However, this issue is under an intense debate, since it has been also found that in the current scenario of the Milky Way, the formation of the inner stellar halo seems to be triggered by the major merger  collision with the Gaia-Enceladus Galaxy \citep{Helmi2018}. In this sense, it has been hypothesized that part of the halo may be associated to the Gaia-Enceladus encounter, and part to an {\sl in situ} halo \citep[e.g.][]{Gallart2019}. It is beyond of the scope of the present work to unravel the origin of the local halo white dwarf population.  Be that as it may,  the formation history found for the oldest white dwarfs of our sample is in agreement with this enhacement of the star formation occurred $\simeq11-12\,$Gyr in the past.

Nevertheless, the hypothesis above mentioned of the origin of these young halo white dwarfs does not exclude other possible scenarios. For instance, that is the case of the well known LP\,93-21 white dwarf, which is also present in our sample as J1045+4509. Firstly discovered in the high-proper motion Luyten Palomar survey \citep{Luyten1968}, this white dwarf has been found to have a hot DQ atmosphere, a mass in the range $1.029-1.10\,M_{\odot}$ and cooling age estimate of $2.28-2.81\,$Gyr \citep[][and references therein]{Kawka2020}. It is worth noting that these values are in perfect in agreement with our estimates of $1.074\pm0.033\,M_{\odot}$ for the mass and $2.69\pm0.25$ for the total age. In a recent paper, \citet{Kawka2020} thoroughly analyze the kinematics of LP\,93-21 reaching the conclusion that this DQ white dwarf is the likely merger product of two other white dwarfs, which rejuvenated leading to a shorter cooling age estimate. Furthermore, the kinematics of LP\,93-21 is compatible with a dwarf galaxy merger event, suggesting that this white dwarf was either captured by the Milky Way or its orbit was affected by the galaxy merger, hence claiming a total age of $\gtrsim 10$\,Gyr. 

On the other hand, we also note that a small fraction of halo white dwarfs could even form as walk-away or runaway secondaries, which are ejected following the disruption of binary systems that underwent either core-collapse or thermonuclear supernova explosions. As an example, \cite{Renzo2019} reported that simulations of systems with primary stars more massive than $7.5\,M_{\odot}$, produce ejected main-sequence stars that will eventually become white dwarfs with velocities in the range of 10--90\,\kms (8\,percent of simulated binaries).
These ejected main-sequence stars would also be rejuvenated due to mass-transfer occurring before the core-collapse supernova of the companion, hence making the future white dwarf appears younger.
Similar or even faster ejection velocities are achieved through the thermonuclear supernova channel \citep[][and refereces therein]{Shen2018}, where the ejected companions could achieve ejection velocities up to 500\,\kms (main-sequence and
helium-core donors) or more than 1\,000\,\kms (white dwarf donors). 

 It is beyond the scope of the present work to ascertain the origin of those young halo white dwarf candidates found here. However, along with the possibility that the age distribution of these objects is compatible with the Sagittarius galaxy encounters, an individual analysis is required to fully unravel their origin.

\subsection{Kinematics of the halo white dwarf sample} 

The stellar parameter analysis done in the previous sections can be complemented with a  kinematic study of the halo white dwarf sample. However, we should be cautious given that only proper motions are provided for the majority of objects of our sample. Consequently, we adopt the standard assumption of null radial velocity in deriving the Galactic components of the velocity $U$, $V$ and $W$. Despite the biases that this assumption may induce, \citep[e.g.][]{Pauli2006,Torres2019} some of the clustering properties of the sample can remain unaltered \citep{Fuchs2011,Torres2019b}. In this sense, we use the integral of motion space as an appropriate tool for analyzing our sample \citep[e.g][]{Helmi2006}. Since the volume of our sample is relatively small in terms of the size of our Galaxy, we can safely approximate the component of angular momentum perpendicular to the Galactic plane, $L_z$, by the $V$-velocity component, as well as the radial and azimuthal Galactic components, $V_R$ and $V_{\phi}$, by the components $U$ and $V$, respectively.

In Figure \ref{f:kin} we show the action space corresponding to the integral of motion $V_{\Delta E}\equiv(U^2+2V^2)^{1/2}$ as a function of $V$ (left panel) and the Toomre diagram (right panel) for the white dwarfs of our halo candidate sample. For helping the analysis we  divided the sample in three groups attending to the total age: young, $T_{\rm age}<8\,$Gyr (yellow dots), middle $8<T_{\rm age}<10\,$Gyr (red dots) and old objects $T_{\rm age}>10\,$Gyr (black dots). Also shown, for illustrative purposes, are the thin (light red dots) and thick disc (light blue dots) velocities for the 100\,pc white dwarf sample from \cite{Torres2019}. The integral of motion $V_{\Delta E}$ can be understood as a measure of the eccentricity, $e$, of the orbit \citep[see][and references therein]{Fuchs2011}. Thus, the left panel shown in Fig. \ref{f:kin} is equivalent to a $(L_z,e)$ diagram. In this sense, we observe that the bulk of old objects of our sample appear with moderately eccentric and retrograde orbits in the range $\approx -100$ to $-200$\,\kms. A value slightly above that of $\approx-220\,$\kms generally adopted for the local standard of rest with respect of the center of the Galaxy, but in agreement for what is considered a classical halo population. On the contrary, those with the highest eccentric and highly retrograde orbits are some of the youngest objects of our sample. These facts suggest to discard these objects as belonging to a typical thick disk population, hence to be thick disk contaminats. Besides, some prograde and relatively middle age objects  appear in our sample. However, we should be cautious about this fact, given that one of the effects of the null radial velocity assumption is the misleading prograde orbits, specially for high speed objects, as pointed out in \cite{Pauli2006}. Regarding the Toomre diagram (right panel of Fig. \ref{f:kin}), a certain clustering seems to be found for the oldest objects of our sample while a greater dispersion is found for the middle and younger ones. In particular, the youngest objects exhibit, on average,  the largest speeds.

\section{Conclusions}
\label{conclusions}

The {\it Gaia} space mission on its DR2 has provided an unprecedented wealth of photometric and astrometric data. In particular, the white dwarf population has found to be nearly complete up to 100\,pc around the Sun. In this sample, 95 white dwarfs have been identified as belonging to a halo population. In the present work, this subsample has been analyzed and the stellar parameters of the objects has been derived. 

Observations made with the Very Large Telescope UT1 equipped with the FORS2 spectrograph allowed us to obtain low-resolution spectra for 27 of our 95 halo white dwarf candidates. By applying a fitting routine to the observed spectra we derive effective temperatures for 24 DCs and 1 DA of the observed sample. We also derived the surface gravity  of the DA white dwarf via fitting its Balmer lines, which results also in obtaining its mass, radius and luminosity via interpolating the observed effective temperature and surface gravity in the appropriate cooling sequences. On the other hand, we apply a procedure based on {\it Gaia} astro-photometry and on a detailed population synthesis code, that permitted to derive the stellar parameters of the white dwarfs. The parameters thus derived were tested with those  available from spectroscopy. The agreement of both methods, in particular in the effective temperature parameter, guarantees us the reliability of our photometric routine to derive the stellar parameters, i.e., mass, radius, bolometric luminosity and age, for the whole sample of 95  halo white dwarf candidates.

The major results found for the different parameter distributions are summarized as follows:
\begin{itemize}
    \item  A generalized version of the modified volume technique, which allow us to take into account all kind of incompleteness sources, has been used to build the 100\,pc halo white dwarf luminosity function. The halo luminosity function reasonably presents the first ever detected evidence of a clear cut-off at faint bolometric magnitudes.  Although the number of objects of our sample is relatively small compared to other published halo white dwarf samples, the completeness analysis reveals $\sim60\%$ completeness for the faint region thus guarantying an acceptable statistical significance of the cut-off found in our sample. The corresponding fitting of the cut-off leads to an age of $\approx12\pm0.5\,$Gyr.
    
    \item The halo white dwarf mass distribution peaks at $0.589\,M_{\odot}$, with most of the stars ($71\%$) having  masses below $0.6\,M_{\odot}$. Only two objects have been found to be more massive than $0.8\,M_{\odot}$. 
    
   \item The majority of white dwarfs ($60\%)$ have total ages (cooling time plus progenitor lifetime) older than $10\,$Gyr. In particular, we found 3 objects with total ages above 12\,Gyr, being the object J1312-4728 the oldest white dwarf found so far with an age of  $12.41\pm0.22\,$Gyr .
   
   \item The star formation history is basically reproduced by a burst of star formation occurring from 10 to 12\,Gyr and extended up to 8 Gyr. The peak of the star formation history is centered at around 11\,Gyr, which is compatible with the current age of the Gaia-Enceladus encounter.
   
   \item $13\%$ of our halo sample is contaminated by high-speed young objects (total age<7\,Gyr). The origin of these white dwarfs is unclear but their age distribution may be compatible with the pass of the Sagittarius galaxy. An individual analysis is required to unravel the origin of each object.
    
    \item Finally, the kinematics analysis of the halo white dwarf sample reveals that there is some clustering of the oldest ($T_{\rm age}>10\,Gyr$) objects of the sample suggesting a common origin, whereas on the contrary, youngest objects exhibit larger eccentric and retrograde orbits. Some prograde orbits have also been found, however, the lack of radial velocity observations prevents us to obtain definitive conclusions.
    
\end{itemize}

In this work we have demonstrated the utility of white dwarfs to address important open questions in astronomy such as the age of the Galactic halo and its star formation history. These objects can also potentially help to reveal the past history and evolution of our Galaxy, an issue that we will analyse in the future with the forthcoming data releases of {\it Gaia}.

\section*{Acknowledgements}

We wish to thank the suggestions and comments of our 
anonymous referee that strongly improved the original version of this paper. This work  was partially supported by the MINECO grant AYA\-2017-86274-P and the Ram\'on y Cajal programme RYC-2016-20254 and by the AGAUR (SGR-661/2017). ST wants to thank Marco Lam for kindly providing the set of luminosity functions used in this work. ST also wants to thank Carme Gallart for useful discussions on the star formation history of the Galaxy. This work has made use of data from the European Space Agency (ESA) mission {\it Gaia} (\url{https://www.cosmos.esa.int/gaia}), processed by the {\it Gaia} Data Processing and Analysis Consortium (DPAC, \url{https://www.cosmos.esa.int/web/gaia/dpac/consortium}). Funding for the DPAC has been provided by national institutions, in particular the institutions participating in the {\it Gaia} Multilateral Agreement.

Based on observations collected at the European Southern Observatory under ESO programme(s) 60.A-9203(E) and 0103.D-0271(A,B,C,D,E).

RR has received funding from the postdoctoral fellowship programme Beatriu de Pin\'os, funded by the Secretary of Universities and Research (Government of Catalonia) and by the Horizon 2020 programme of research and innovation of the European Union under the Maria Sk\l{}odowska-Curie grant agreement No 801370.

\section*{Data Availability Statement}
The data underlying this article are available in the article.  Supplementary material will be shared on reasonable request to the corresponding author.



\bibliographystyle{mnras}
\bibliography{hwlf}

\begin{thebibliography}{}
\makeatletter
\relax
\def\mn@urlcharsother{\let\do\@makeother \do\$\do\&\do\#\do\^\do\_\do\%\do\~}
\def\mn@doi{\begingroup\mn@urlcharsother \@ifnextchar [ {\mn@doi@}
  {\mn@doi@[]}}
\def\mn@doi@[#1]#2{\def\@tempa{#1}\ifx\@tempa\@empty \href
  {http://dx.doi.org/#2} {doi:#2}\else \href {http://dx.doi.org/#2} {#1}\fi
  \endgroup}
\def\mn@eprint#1#2{\mn@eprint@#1:#2::\@nil}
\def\mn@eprint@arXiv#1{\href {http://arxiv.org/abs/#1} {{\tt arXiv:#1}}}
\def\mn@eprint@dblp#1{\href {http://dblp.uni-trier.de/rec/bibtex/#1.xml}
  {dblp:#1}}
\def\mn@eprint@#1:#2:#3:#4\@nil{\def\@tempa {#1}\def\@tempb {#2}\def\@tempc
  {#3}\ifx \@tempc \@empty \let \@tempc \@tempb \let \@tempb \@tempa \fi \ifx
  \@tempb \@empty \def\@tempb {arXiv}\fi \@ifundefined
  {mn@eprint@\@tempb}{\@tempb:\@tempc}{\expandafter \expandafter \csname
  mn@eprint@\@tempb\endcsname \expandafter{\@tempc}}}

\bibitem[\protect\citeauthoryear{Alcock et~al.,}{Alcock
  et~al.}{2000}]{Alcock2000}
Alcock C.,  et~al., 2000, The Astrophysical Journal, 542, 281

\bibitem[\protect\citeauthoryear{{Althaus}, {C{\'o}rsico}, {Isern}  \&
  {Garc{\'{\i}}a-Berro}}{{Althaus} et~al.}{2010}]{Althaus2010}
{Althaus} L.~G.,  {C{\'o}rsico} A.~H.,  {Isern} J.,   {Garc{\'{\i}}a-Berro} E.,
   2010, \mn@doi [\aapr] {10.1007/s00159-010-0033-1}, \href
  {http://cdsads.u-strasbg.fr/abs/2010A%26ARv..18..471A} {18, 471}

\bibitem[\protect\citeauthoryear{{Althaus}, {Camisassa}, {Miller Bertolami},
  {C{\'o}rsico}  \& {Garc{\'\i}a-Berro}}{{Althaus} et~al.}{2015}]{Althaus2015}
{Althaus} L.~G.,  {Camisassa} M.~E.,  {Miller Bertolami} M.~M.,  {C{\'o}rsico}
  A.~H.,   {Garc{\'\i}a-Berro} E.,  2015, \mn@doi [\aap]
  {10.1051/0004-6361/201424922}, \href
  {https://ui.adsabs.harvard.edu/abs/2015A&A...576A...9A} {576, A9}

\bibitem[\protect\citeauthoryear{{Appenzeller} et~al.,}{{Appenzeller}
  et~al.}{1998}]{Appenzeller1998}
{Appenzeller} I.,  et~al., 1998, The Messenger, \href
  {https://ui.adsabs.harvard.edu/abs/1998Msngr..94....1A} {94, 1}

\bibitem[\protect\citeauthoryear{{Bergeron}}{{Bergeron}}{2001}]{Bergeron2001}
{Bergeron} P.,  2001, \mn@doi [\apj] {10.1086/322316}, \href
  {https://ui.adsabs.harvard.edu/abs/2001ApJ...558..369B} {558, 369}

\bibitem[\protect\citeauthoryear{{Bergeron}, {Ruiz}, {Hamuy}, {Leggett},
  {Currie}, {Lajoie}  \& {Dufour}}{{Bergeron} et~al.}{2005}]{Bergeron2005}
{Bergeron} P.,  {Ruiz} M.~T.,  {Hamuy} M.,  {Leggett} S.~K.,  {Currie} M.~J.,
  {Lajoie} C.-P.,   {Dufour} P.,  2005, \mn@doi [\apj] {10.1086/429715}, \href
  {http://adsabs.harvard.edu/abs/2005ApJ...625..838B} {625, 838}

\bibitem[\protect\citeauthoryear{{Bergeron}, {Dufour}, {Fontaine}, {Coutu},
  {Blouin}, {Genest-Beaulieu}, {B{\'e}dard}  \& {Rolland }}{{Bergeron}
  et~al.}{2019}]{Bergeron2019}
{Bergeron} P.,  {Dufour} P.,  {Fontaine} G.,  {Coutu} S.,  {Blouin} S.,
  {Genest-Beaulieu} C.,  {B{\'e}dard} A.,   {Rolland } B.,  2019, \mn@doi
  [\apj] {10.3847/1538-4357/ab153a}, \href
  {https://ui.adsabs.harvard.edu/abs/2019ApJ...876...67B} {876, 67}

\bibitem[\protect\citeauthoryear{{Calamida} et~al.,}{{Calamida}
  et~al.}{2014}]{Calamida2014}
{Calamida} A.,  et~al., 2014, \mn@doi [\apj] {10.1088/0004-637X/790/2/164},
  \href {http://adsabs.harvard.edu/abs/2014ApJ...790..164C} {790, 164}

\bibitem[\protect\citeauthoryear{{Camisassa}, {Althaus}, {C{\'o}rsico},
  {Vinyoles}, {Serenelli}, {Isern}, {Miller Bertolami}  \&
  {Garc{\'{\i}}a-Berro}}{{Camisassa} et~al.}{2016}]{Camisassa2016}
{Camisassa} M.~E.,  {Althaus} L.~G.,  {C{\'o}rsico} A.~H.,  {Vinyoles} N.,
  {Serenelli} A.~M.,  {Isern} J.,  {Miller Bertolami} M.~M.,
  {Garc{\'{\i}}a-Berro} E.,  2016, \mn@doi [\apj]
  {10.3847/0004-637X/823/2/158}, \href
  {http://adsabs.harvard.edu/abs/2016ApJ...823..158C} {823, 158}

\bibitem[\protect\citeauthoryear{{Camisassa}, {Althaus}, {Rohrmann},
  {Garc{\'{\i}}a-Berro}, {Torres}, {C{\'o}rsico}  \& {Wachlin}}{{Camisassa}
  et~al.}{2017}]{Camisassa2017}
{Camisassa} M.~E.,  {Althaus} L.~G.,  {Rohrmann} R.~D.,  {Garc{\'{\i}}a-Berro}
  E.,  {Torres} S.,  {C{\'o}rsico} A.~H.,   {Wachlin} F.~C.,  2017, \mn@doi
  [\apj] {10.3847/1538-4357/aa6797}, \href
  {http://adsabs.harvard.edu/abs/2017ApJ...839...11C} {839, 11}

\bibitem[\protect\citeauthoryear{{Camisassa} et~al.,}{{Camisassa}
  et~al.}{2019}]{Camisassa2019}
{Camisassa} M.~E.,  et~al., 2019, \mn@doi [\aap] {10.1051/0004-6361/201833822},
  \href {https://ui.adsabs.harvard.edu/abs/2019A&A...625A..87C} {625, A87}

\bibitem[\protect\citeauthoryear{{Catal{\'a}n}, {Isern}, {Garc{\'\i}a-Berro},
  {Ribas}, {Allende Prieto}  \& {Bonanos}}{{Catal{\'a}n}
  et~al.}{2008}]{Catalan2008}
{Catal{\'a}n} S.,  {Isern} J.,  {Garc{\'\i}a-Berro} E.,  {Ribas} I.,  {Allende
  Prieto} C.,   {Bonanos} A.~Z.,  2008, \mn@doi [\aap]
  {10.1051/0004-6361:20078111}, \href
  {https://ui.adsabs.harvard.edu/abs/2008A&A...477..213C} {477, 213}

\bibitem[\protect\citeauthoryear{{Catal{\'a}n} et~al.,}{{Catal{\'a}n}
  et~al.}{2012}]{Catalan2012}
{Catal{\'a}n} S.,  et~al., 2012, \mn@doi [\aap] {10.1051/0004-6361/201220191},
  \href {https://ui.adsabs.harvard.edu/abs/2012A&A...546L...3C} {546, L3}

\bibitem[\protect\citeauthoryear{{Cojocaru}, {Torres}, {Althaus}, {Isern}  \&
  {Garc{\'{\i}}a-Berro}}{{Cojocaru} et~al.}{2015}]{Cojocaru2015}
{Cojocaru} R.,  {Torres} S.,  {Althaus} L.~G.,  {Isern} J.,
  {Garc{\'{\i}}a-Berro} E.,  2015, \mn@doi [\aap]
  {10.1051/0004-6361/201526550}, \href
  {http://adsabs.harvard.edu/abs/2015A%26A...581A.108C} {581, A108}

\bibitem[\protect\citeauthoryear{{Felten}}{{Felten}}{1976}]{Felten1976}
{Felten} J.~E.,  1976, \mn@doi [\apj] {10.1086/154538}, \href
  {https://ui.adsabs.harvard.edu/abs/1976ApJ...207..700F} {207, 700}

\bibitem[\protect\citeauthoryear{Flynn, Holopainen  \& Holmberg}{Flynn
  et~al.}{2003}]{Flynn2003}
Flynn C.,  Holopainen J.,   Holmberg J.,  2003, Monthly Notices of the Royal
  Astronomical Society, 339, 817

\bibitem[\protect\citeauthoryear{{Fuchs} \& {Dettbarn}}{{Fuchs} \&
  {Dettbarn}}{2011}]{Fuchs2011}
{Fuchs} B.,  {Dettbarn} C.,  2011, \mn@doi [\aj] {10.1088/0004-6256/141/1/5},
  \href {https://ui.adsabs.harvard.edu/abs/2011AJ....141....5F} {141, 5}

\bibitem[\protect\citeauthoryear{{Gaia Collaboration} et~al.,}{{Gaia
  Collaboration} et~al.}{2018}]{Gaiacol2018}
{Gaia Collaboration} et~al., 2018, \mn@doi [\aap]
  {10.1051/0004-6361/201833051}, \href
  {https://ui.adsabs.harvard.edu/abs/2018A&A...616A...1G} {616, A1}

\bibitem[\protect\citeauthoryear{{Gaia Collaboration} et~al.,}{{Gaia
  Collaboration} et~al.}{2020}]{Gaiacol2020}
{Gaia Collaboration} et~al., 2020, arXiv e-prints, \href
  {https://ui.adsabs.harvard.edu/abs/2020arXiv201202061G} {p. arXiv:2012.02061}

\bibitem[\protect\citeauthoryear{{Gallart}, {Bernard}, {Brook}, {Ruiz-Lara},
  {Cassisi}, {Hill}  \& {Monelli}}{{Gallart} et~al.}{2019}]{Gallart2019}
{Gallart} C.,  {Bernard} E.~J.,  {Brook} C.~B.,  {Ruiz-Lara} T.,  {Cassisi} S.,
   {Hill} V.,   {Monelli} M.,  2019, \mn@doi [Nature Astronomy]
  {10.1038/s41550-019-0829-5}, \href
  {https://ui.adsabs.harvard.edu/abs/2019NatAs...3..932G} {3, 932}

\bibitem[\protect\citeauthoryear{{Garc{\'{\i}}a-Berro} \&
  {Oswalt}}{{Garc{\'{\i}}a-Berro} \& {Oswalt}}{2016}]{GBerroOswalt2016}
{Garc{\'{\i}}a-Berro} E.,  {Oswalt} T.~D.,  2016, \mn@doi [\nar]
  {10.1016/j.newar.2016.08.001}, \href
  {http://adsabs.harvard.edu/abs/2016NewAR..72....1G} {72, 1}

\bibitem[\protect\citeauthoryear{{Garc\'{i}a-Berro}, E., {Isern}  \&
  {Burkert}}{{Garc\'{i}a-Berro} et~al.}{1999}]{GBerro1999}
{Garc\'{i}a-Berro} E. {Torres} S.,  {Isern} J.,   {Burkert} A.,  1999, \mnras,
  302, 173

\bibitem[\protect\citeauthoryear{{Garc{\'{\i}}a-Berro}, {Torres}, {Isern}  \&
  {Burkert}}{{Garc{\'{\i}}a-Berro} et~al.}{2004}]{GBerro2004}
{Garc{\'{\i}}a-Berro} E.,  {Torres} S.,  {Isern} J.,   {Burkert} A.,  2004,
  \mn@doi [\aap] {10.1051/0004-6361:20034541}, \href
  {http://adsabs.harvard.edu/abs/2004A%26A...418...53G} {418, 53}

\bibitem[\protect\citeauthoryear{{Garc{\'{\i}}a-Berro}
  et~al.,}{{Garc{\'{\i}}a-Berro} et~al.}{2010}]{GBerro2010}
{Garc{\'{\i}}a-Berro} E.,  et~al., 2010, \mn@doi [\nat] {10.1038/nature09045},
  \href {http://adsabs.harvard.edu/abs/2010Natur.465..194G} {465, 194}

\bibitem[\protect\citeauthoryear{{Geijo}, {Torres}, {Isern}  \&
  {Garc{\'\i}a-Berro}}{{Geijo} et~al.}{2006}]{Geijo2006}
{Geijo} E.~M.,  {Torres} S.,  {Isern} J.,   {Garc{\'\i}a-Berro} E.,  2006,
  \mn@doi [\mnras] {10.1111/j.1365-2966.2006.10354.x}, \href
  {https://ui.adsabs.harvard.edu/abs/2006MNRAS.369.1654G} {369, 1654}

\bibitem[\protect\citeauthoryear{{Gentile Fusillo} et~al.,}{{Gentile Fusillo}
  et~al.}{2019}]{Fusillo2019}
{Gentile Fusillo} N.~P.,  et~al., 2019, \mn@doi [\mnras]
  {10.1093/mnras/sty3016}, \href
  {https://ui.adsabs.harvard.edu/abs/2019MNRAS.482.4570G} {482, 4570}

\bibitem[\protect\citeauthoryear{{Gianninas}, {Curd}, {Thorstensen}, {Kilic},
  {Bergeron}, {Andrews}, {Canton}  \& {Ag{\"u}eros}}{{Gianninas}
  et~al.}{2015}]{Gianninas2015}
{Gianninas} A.,  {Curd} B.,  {Thorstensen} J.~R.,  {Kilic} M.,  {Bergeron} P.,
  {Andrews} J.~J.,  {Canton} P.,   {Ag{\"u}eros} M.~A.,  2015, \mn@doi [\mnras]
  {10.1093/mnras/stv545}, \href
  {http://adsabs.harvard.edu/abs/2015MNRAS.449.3966G} {449, 3966}

\bibitem[\protect\citeauthoryear{{Hall}, {Kowalski}, {Harris}, {Awal},
  {Leggett}, {Kilic}, {Anderson}  \& {Gates}}{{Hall} et~al.}{2008}]{Hall2008}
{Hall} P.~B.,  {Kowalski} P.~M.,  {Harris} H.~C.,  {Awal} A.,  {Leggett} S.~K.,
   {Kilic} M.,  {Anderson} S.~F.,   {Gates} E.,  2008, \mn@doi [\aj]
  {10.1088/0004-6256/136/1/76}, \href
  {https://ui.adsabs.harvard.edu/abs/2008AJ....136...76H} {136, 76}

\bibitem[\protect\citeauthoryear{{Hambly}, {Smartt}  \& {Hodgkin}}{{Hambly}
  et~al.}{1997}]{Hambly1997}
{Hambly} N.~C.,  {Smartt} S.~J.,   {Hodgkin} S.~T.,  1997, \mn@doi [\apjl]
  {10.1086/316797}, \href {http://adsabs.harvard.edu/abs/1997ApJ...489L.157H}
  {489, L157}

\bibitem[\protect\citeauthoryear{{Hansen} et~al.,}{{Hansen}
  et~al.}{2013}]{Hansen2013}
{Hansen} B.~M.~S.,  et~al., 2013, \mn@doi [\nat] {10.1038/nature12334}, \href
  {http://adsabs.harvard.edu/abs/2013Natur.500...51H} {500, 51}

\bibitem[\protect\citeauthoryear{Helmi, Navarro, Nordström, Holmberg, Abadi
  \& Steinmetz}{Helmi et~al.}{2006}]{Helmi2006}
Helmi A.,  Navarro J.~F.,  Nordström B.,  Holmberg J.,  Abadi M.~G.,
  Steinmetz M.,  2006, \mn@doi [Monthly Notices of the Royal Astronomical
  Society] {10.1111/j.1365-2966.2005.09818.x}, 365, 1309

\bibitem[\protect\citeauthoryear{{Helmi}, {Babusiaux}, {Koppelman}, {Massari},
  {Veljanoski}  \& {Brown}}{{Helmi} et~al.}{2018}]{Helmi2018}
{Helmi} A.,  {Babusiaux} C.,  {Koppelman} H.~H.,  {Massari} D.,  {Veljanoski}
  J.,   {Brown} A. G.~A.,  2018, \mn@doi [\nat] {10.1038/s41586-018-0625-x},
  \href {https://ui.adsabs.harvard.edu/abs/2018Natur.563...85H} {563, 85}

\bibitem[\protect\citeauthoryear{Hidalgo et~al.,}{Hidalgo
  et~al.}{2018}]{Hidalgo2018}
Hidalgo S.,  et~al., 2018, \mn@doi [The Astrophysical Journal]
  {10.3847/1538-4357/aab158}, 856

\bibitem[\protect\citeauthoryear{{Ibata}, {Irwin}, {Bienaym{\'e}}, {Scholz}  \&
  {Guibert}}{{Ibata} et~al.}{2000}]{Ibata2000}
{Ibata} R.,  {Irwin} M.,  {Bienaym{\'e}} O.,  {Scholz} R.,   {Guibert} J.,
  2000, \mn@doi [\apjl] {10.1086/312566}, \href
  {http://adsabs.harvard.edu/abs/2000ApJ...532L..41I} {532, L41}

\bibitem[\protect\citeauthoryear{{Isern}, {Garc{\'{\i}}a-Berro}, {Hernanz},
  {Mochkovitch}  \& {Torres}}{{Isern} et~al.}{1998}]{Isern1998}
{Isern} J.,  {Garc{\'{\i}}a-Berro} E.,  {Hernanz} M.,  {Mochkovitch} R.,
  {Torres} S.,  1998, \mn@doi [\apj] {10.1086/305977}, \href
  {http://adsabs.harvard.edu/abs/1998ApJ...503..239I} {503, 239}

\bibitem[\protect\citeauthoryear{{Jeffery}, {von Hippel}, {DeGennaro}, {van
  Dyk}, {Stein}  \& {Jefferys}}{{Jeffery} et~al.}{2011}]{Jeffery2011}
{Jeffery} E.~J.,  {von Hippel} T.,  {DeGennaro} S.,  {van Dyk} D.~A.,  {Stein}
  N.,   {Jefferys} W.~H.,  2011, \mn@doi [\apj] {10.1088/0004-637X/730/1/35},
  \href {http://adsabs.harvard.edu/abs/2011ApJ...730...35J} {730, 35}

\bibitem[\protect\citeauthoryear{{Jim{\'e}nez-Esteban}, {Torres},
  {Rebassa-Mansergas}, {Skorobogatov}, {Solano}, {Cantero}  \&
  {Rodrigo}}{{Jim{\'e}nez-Esteban} et~al.}{2018}]{Jimenez2018}
{Jim{\'e}nez-Esteban} F.~M.,  {Torres} S.,  {Rebassa-Mansergas} A.,
  {Skorobogatov} G.,  {Solano} E.,  {Cantero} C.,   {Rodrigo} C.,  2018,
  \mn@doi [\mnras] {10.1093/mnras/sty2120}, \href
  {http://adsabs.harvard.edu/abs/2018MNRAS.480.4505J} {480, 4505}

\bibitem[\protect\citeauthoryear{Kalirai}{Kalirai}{2012}]{Kalirai2012}
Kalirai J.~S.,  2012, \mn@doi [Nature] {10.1038/nature11062}, \href
  {https://doi.org/10.1038/nature11062} {486, 90}

\bibitem[\protect\citeauthoryear{Kawka \& Vennes}{Kawka \&
  Vennes}{2012}]{Kawka2012}
Kawka A.,  Vennes S.,  2012, \mn@doi [Monthly Notices of the Royal Astronomical
  Society] {10.1111/j.1365-2966.2012.21574.x}, 425, 1394

\bibitem[\protect\citeauthoryear{{Kawka}, {Vennes}  \& {Ferrario}}{{Kawka}
  et~al.}{2020}]{Kawka2020}
{Kawka} A.,  {Vennes} S.,   {Ferrario} L.,  2020, \mn@doi [\mnras]
  {10.1093/mnrasl/slz165}, \href
  {https://ui.adsabs.harvard.edu/abs/2020MNRAS.491L..40K} {491, L40}

\bibitem[\protect\citeauthoryear{{Kilic}, {von Hippel}, {Mendez}  \&
  {Winget}}{{Kilic} et~al.}{2004}]{Kilic2004}
{Kilic} M.,  {von Hippel} T.,  {Mendez} R.~A.,   {Winget} D.~E.,  2004, \mn@doi
  [\apj] {10.1086/421343}, \href
  {http://adsabs.harvard.edu/abs/2004ApJ...609..766K} {609, 766}

\bibitem[\protect\citeauthoryear{{Kilic}, {Thorstensen}, {Kowalski}  \&
  {Andrews}}{{Kilic} et~al.}{2012}]{Kilic2012}
{Kilic} M.,  {Thorstensen} J.~R.,  {Kowalski} P.~M.,   {Andrews} J.,  2012,
  \mn@doi [\mnras] {10.1111/j.1745-3933.2012.01271.x}, \href
  {https://ui.adsabs.harvard.edu/abs/2012MNRAS.423L.132K} {423, L132}

\bibitem[\protect\citeauthoryear{{Kilic}, {Munn}, {Harris}, {von Hippel},
  {Liebert}, {Williams}, {Jeffery}  \& {DeGennaro}}{{Kilic}
  et~al.}{2017}]{Kilic2017}
{Kilic} M.,  {Munn} J.~A.,  {Harris} H.~C.,  {von Hippel} T.,  {Liebert} J.~W.,
   {Williams} K.~A.,  {Jeffery} E.,   {DeGennaro} S.,  2017, \mn@doi [\apj]
  {10.3847/1538-4357/aa62a5}, \href
  {http://adsabs.harvard.edu/abs/2017ApJ...837..162K} {837, 162}

\bibitem[\protect\citeauthoryear{{Kilic}, {Bergeron}, {Dame}, {Hambly},
  {Rowell}  \& {Crawford}}{{Kilic} et~al.}{2018}]{Kilic2018}
{Kilic} M.,  {Bergeron} P.,  {Dame} K.,  {Hambly} N.~C.,  {Rowell} N.,
  {Crawford} C.~L.,  2018, \mn@doi [\mnras] {10.1093/mnras/sty2755}, \href
  {http://adsabs.harvard.edu/abs/2018MNRAS.tmp.2629K} {}

\bibitem[\protect\citeauthoryear{{Koester}}{{Koester}}{2010}]{Koester10}
{Koester} D.,  2010, \memsai, \href
  {http://adsabs.harvard.edu/abs/2010MmSAI..81..921K} {81, 921}

\bibitem[\protect\citeauthoryear{Lam, Rowell  \& Hambly}{Lam
  et~al.}{2015}]{Lam2015}
Lam M.~C.,  Rowell N.,   Hambly N.~C.,  2015, \mn@doi [Monthly Notices of the
  Royal Astronomical Society] {10.1093/mnras/stv876}, 450, 4098

\bibitem[\protect\citeauthoryear{Lam et~al.,}{Lam et~al.}{2018}]{Lam2018}
Lam M.~C.,  et~al., 2018, \mn@doi [Monthly Notices of the Royal Astronomical
  Society] {10.1093/mnras/sty2710}, 482, 715

\bibitem[\protect\citeauthoryear{{Liebert}, {Dahn}  \& {Monet}}{{Liebert}
  et~al.}{1989}]{Liebert1989}
{Liebert} J.,  {Dahn} C.~C.,   {Monet} D.~G.,  1989, in {Wegner} G.,  ed.,
  Lecture Notes in Physics, Berlin Springer Verlag Vol. 328, IAU Colloq. 114:
  White Dwarfs. pp 15--23, \mn@doi{10.1007/3-540-51031-1_287}

\bibitem[\protect\citeauthoryear{{Luyten}}{{Luyten}}{1968}]{Luyten1968}
{Luyten} W.~J.,  1968, Proper Motion Survey, University of Minnesota, \href
  {https://ui.adsabs.harvard.edu/abs/1968PMMin..13....1L} {13, 1}

\bibitem[\protect\citeauthoryear{{Marsh}}{{Marsh}}{1989}]{Marsh1989}
{Marsh} T.~R.,  1989, \mn@doi [\pasp] {10.1086/132570}, \href
  {https://ui.adsabs.harvard.edu/abs/1989PASP..101.1032M} {101, 1032}

\bibitem[\protect\citeauthoryear{{McCleery} et~al.,}{{McCleery}
  et~al.}{2020}]{McCleery2020}
{McCleery} J.,  et~al., 2020, \mn@doi [\mnras] {10.1093/mnras/staa2030}, \href
  {https://ui.adsabs.harvard.edu/abs/2020MNRAS.499.1890M} {499, 1890}

\bibitem[\protect\citeauthoryear{{Mochkovitch}, {Garcia-Berro}, {Hernanz},
  {Isern}  \& {Panis}}{{Mochkovitch} et~al.}{1990}]{Mochkovitch1990}
{Mochkovitch} R.,  {Garcia-Berro} E.,  {Hernanz} M.,  {Isern} J.,   {Panis}
  J.~F.,  1990, \aap, \href
  {https://ui.adsabs.harvard.edu/abs/1990A&A...233..456M} {233, 456}

\bibitem[\protect\citeauthoryear{{Monet}, {Fisher}, {Liebert}, {Canzian},
  {Harris}  \& {Reid}}{{Monet} et~al.}{2000}]{Monet2000}
{Monet} D.~G.,  {Fisher} M.~D.,  {Liebert} J.,  {Canzian} B.,  {Harris} H.~C.,
   {Reid} I.~N.,  2000, \mn@doi [\aj] {10.1086/301530}, \href
  {https://ui.adsabs.harvard.edu/abs/2000AJ....120.1541M} {120, 1541}

\bibitem[\protect\citeauthoryear{Munn et~al.,}{Munn et~al.}{2016}]{Munn2016}
Munn J.~A.,  et~al., 2016, \mn@doi [The Astronomical Journal]
  {10.3847/1538-3881/153/1/10}, 153, 10

\bibitem[\protect\citeauthoryear{{Napiwotzki} et~al.,}{{Napiwotzki}
  et~al.}{2020}]{Napiwotzki2020}
{Napiwotzki} R.,  et~al., 2020, VizieR Online Data Catalog, \href
  {https://ui.adsabs.harvard.edu/abs/2020yCat..36380131N} {pp J/A+A/638/A131}

\bibitem[\protect\citeauthoryear{{Nelson}, {Cook}, {Axelrod}, {Mould}  \&
  {Alcock}}{{Nelson} et~al.}{2002}]{Nelson2002}
{Nelson} C.~A.,  {Cook} K.~H.,  {Axelrod} T.~S.,  {Mould} J.~R.,   {Alcock} C.,
   2002, \mn@doi [\apj] {10.1086/340759}, \href
  {https://ui.adsabs.harvard.edu/abs/2002ApJ...573..644N} {573, 644}

\bibitem[\protect\citeauthoryear{{Oppenheimer} et~al.,}{{Oppenheimer}
  et~al.}{2001}]{Oppenheimer2001}
{Oppenheimer} B.~R.,  et~al., 2001, \mn@doi [\apj] {10.1086/319718}, \href
  {http://adsabs.harvard.edu/abs/2001ApJ...550..448O} {550, 448}

\bibitem[\protect\citeauthoryear{{Pauli}, {Napiwotzki}, {Heber}, {Altmann}  \&
  {Odenkirchen}}{{Pauli} et~al.}{2006}]{Pauli2006}
{Pauli} E.-M.,  {Napiwotzki} R.,  {Heber} U.,  {Altmann} M.,   {Odenkirchen}
  M.,  2006, \mn@doi [\aap] {10.1051/0004-6361:20052730}, 447, 173

\bibitem[\protect\citeauthoryear{{Planck Collaboration} et~al.,}{{Planck
  Collaboration} et~al.}{2016}]{PlanckCol2016}
{Planck Collaboration} et~al., 2016, \mn@doi [\aap]
  {10.1051/0004-6361/201525830}, \href
  {https://ui.adsabs.harvard.edu/abs/2016A&A...594A..13P} {594, A13}

\bibitem[\protect\citeauthoryear{{Rebassa-Mansergas}, {G{\"a}nsicke},
  {Rodr{\'\i}guez-Gil}, {Schreiber}  \& {Koester}}{{Rebassa-Mansergas}
  et~al.}{2007}]{Rebassa2007}
{Rebassa-Mansergas} A.,  {G{\"a}nsicke} B.~T.,  {Rodr{\'\i}guez-Gil} P.,
  {Schreiber} M.~R.,   {Koester} D.,  2007, \mn@doi [\mnras]
  {10.1111/j.1365-2966.2007.12288.x}, \href
  {https://ui.adsabs.harvard.edu/abs/2007MNRAS.382.1377R} {382, 1377}

\bibitem[\protect\citeauthoryear{{Renzo} et~al.,}{{Renzo}
  et~al.}{2019}]{Renzo2019}
{Renzo} M.,  et~al., 2019, \mn@doi [\aap] {10.1051/0004-6361/201833297}, \href
  {https://ui.adsabs.harvard.edu/abs/2019A&A...624A..66R} {624, A66}

\bibitem[\protect\citeauthoryear{{Rowell} \& {Hambly}}{{Rowell} \&
  {Hambly}}{2011}]{Rowell2011}
{Rowell} N.,  {Hambly} N.~C.,  2011, \mn@doi [\mnras]
  {10.1111/j.1365-2966.2011.18976.x}, \href
  {http://cdsads.u-strasbg.fr/abs/2011MNRAS.417...93R} {417, 93}

\bibitem[\protect\citeauthoryear{{Ruiz-Lara}, {Gallart}, {Bernard}  \&
  {Cassisi}}{{Ruiz-Lara} et~al.}{2020}]{RuizLara2020}
{Ruiz-Lara} T.,  {Gallart} C.,  {Bernard} E.~J.,   {Cassisi} S.,  2020, \mn@doi
  [Nature Astronomy] {10.1038/s41550-020-1097-0}, \href
  {https://ui.adsabs.harvard.edu/abs/2020NatAs.tmp..111R} {}

\bibitem[\protect\citeauthoryear{{Schmidt}}{{Schmidt}}{1968}]{Schmidt1968}
{Schmidt} M.,  1968, \mn@doi [\apj] {10.1086/149446}, \href
  {https://ui.adsabs.harvard.edu/abs/1968ApJ...151..393S} {151, 393}

\bibitem[\protect\citeauthoryear{{Shen} et~al.,}{{Shen}
  et~al.}{2018}]{Shen2018}
{Shen} K.~J.,  et~al., 2018, \mn@doi [\apj] {10.3847/1538-4357/aad55b}, \href
  {https://ui.adsabs.harvard.edu/abs/2018ApJ...865...15S} {865, 15}

\bibitem[\protect\citeauthoryear{{Si}, {van Dyk}, {von Hippel}, {Robinson},
  {Webster}  \& {Stenning}}{{Si} et~al.}{2017}]{Si2017}
{Si} S.,  {van Dyk} D.~A.,  {von Hippel} T.,  {Robinson} E.,  {Webster} A.,
  {Stenning} D.,  2017, \mn@doi [\mnras] {10.1093/mnras/stx765}, \href
  {http://adsabs.harvard.edu/abs/2017MNRAS.468.4374S} {468, 4374}

\bibitem[\protect\citeauthoryear{{Torres} \& {Garc{\'{\i}}a-Berro}}{{Torres} \&
  {Garc{\'{\i}}a-Berro}}{2016}]{Torres2016}
{Torres} S.,  {Garc{\'{\i}}a-Berro} E.,  2016, \mn@doi [\aap]
  {10.1051/0004-6361/201528059}, \href
  {http://adsabs.harvard.edu/abs/2016A%26A...588A..35T} {588, A35}

\bibitem[\protect\citeauthoryear{Torres, Garc{\'{\i}}a-Berro  \& Isern}{Torres
  et~al.}{1998}]{Torres1998}
Torres S.,  Garc{\'{\i}}a-Berro E.,   Isern J.,  1998, The Astrophysical
  Journal Letters, 508, L71

\bibitem[\protect\citeauthoryear{Torres, Garc{\'{\i}}a-Berro, Burkert  \&
  Isern}{Torres et~al.}{2001}]{Torres2001}
Torres S.,  Garc{\'{\i}}a-Berro E.,  Burkert A.,   Isern J.,  2001, \mn@doi
  [\mnras] {10.1046/j.1365-8711.2001.04885.x}, 328, 492

\bibitem[\protect\citeauthoryear{{Torres}, {Garc{\'{\i}}a-Berro}, {Burkert}  \&
  {Isern}}{{Torres} et~al.}{2002}]{Torres2002}
{Torres} S.,  {Garc{\'{\i}}a-Berro} E.,  {Burkert} A.,   {Isern} J.,  2002,
  \mn@doi [\mnras] {10.1046/j.1365-8711.2002.05830.x}, \href
  {http://adsabs.harvard.edu/abs/2002MNRAS.336..971T} {336, 971}

\bibitem[\protect\citeauthoryear{{Torres}, {Garc{\'{\i}}a-Berro}, {Isern}  \&
  {Figueras}}{{Torres} et~al.}{2005}]{Torres2005}
{Torres} S.,  {Garc{\'{\i}}a-Berro} E.,  {Isern} J.,   {Figueras} F.,  2005,
  \mn@doi [\mnras] {10.1111/j.1365-2966.2005.09128.x}, \href
  {http://adsabs.harvard.edu/abs/2005MNRAS.360.1381T} {360, 1381}

\bibitem[\protect\citeauthoryear{{Torres}, {Garc{\'{\i}}a-Berro}, {Althaus}  \&
  {Camisassa}}{{Torres} et~al.}{2015}]{Torres2015}
{Torres} S.,  {Garc{\'{\i}}a-Berro} E.,  {Althaus} L.~G.,   {Camisassa} M.~E.,
  2015, \mn@doi [\aap] {10.1051/0004-6361/201526157}, \href
  {http://adsabs.harvard.edu/abs/2015A%26A...581A..90T} {581, A90}

\bibitem[\protect\citeauthoryear{{Torres}, {Garc{\'{\i}}a-Berro}, {Cojocaru}
  \& {Calamida}}{{Torres} et~al.}{2018}]{Torres2018}
{Torres} S.,  {Garc{\'{\i}}a-Berro} E.,  {Cojocaru} R.,   {Calamida} A.,  2018,
  \mn@doi [\mnras] {10.1093/mnras/sty289}, \href
  {http://adsabs.harvard.edu/abs/2018MNRAS.tmp..283T} {}

\bibitem[\protect\citeauthoryear{{Torres}, {Cantero}, {Rebassa-Mansergas},
  {Skorobogatov}, {Jim{\'e}nez-Esteban}  \& {Solano}}{{Torres}
  et~al.}{2019a}]{Torres2019}
{Torres} S.,  {Cantero} C.,  {Rebassa-Mansergas} A.,  {Skorobogatov} G.,
  {Jim{\'e}nez-Esteban} F.~M.,   {Solano} E.,  2019a, \mn@doi [\mnras]
  {10.1093/mnras/stz814}, \href
  {https://ui.adsabs.harvard.edu/abs/2019MNRAS.485.5573T} {485, 5573}

\bibitem[\protect\citeauthoryear{{Torres}, {Cantero}, {Camisassa}, {Antoja},
  {Rebassa-Mansergas}, {Althaus}, {Thelemaque}  \& {C{\'a}novas}}{{Torres}
  et~al.}{2019b}]{Torres2019b}
{Torres} S.,  {Cantero} C.,  {Camisassa} M.~E.,  {Antoja} T.,
  {Rebassa-Mansergas} A.,  {Althaus} L. r.~G.,  {Thelemaque} T.,
  {C{\'a}novas} H.,  2019b, \mn@doi [\aap] {10.1051/0004-6361/201936244}, \href
  {https://ui.adsabs.harvard.edu/abs/2019A&A...629L...6T} {629, L6}

\bibitem[\protect\citeauthoryear{{Tremblay}, {Cummings}, {Kalirai},
  {G{\"a}nsicke}, {Gentile-Fusillo}  \& {Raddi}}{{Tremblay}
  et~al.}{2016}]{Tremblay2016}
{Tremblay} P.~E.,  {Cummings} J.,  {Kalirai} J.~S.,  {G{\"a}nsicke} B.~T.,
  {Gentile-Fusillo} N.,   {Raddi} R.,  2016, \mn@doi [\mnras]
  {10.1093/mnras/stw1447}, \href
  {https://ui.adsabs.harvard.edu/abs/2016MNRAS.461.2100T} {461, 2100}

\bibitem[\protect\citeauthoryear{{Tremblay} et~al.,}{{Tremblay}
  et~al.}{2020}]{Tremblay2020}
{Tremblay} P.~E.,  et~al., 2020, \mn@doi [\mnras] {10.1093/mnras/staa1892},
  \href {https://ui.adsabs.harvard.edu/abs/2020MNRAS.tmp.2021T} {}

\makeatother
\end{thebibliography}


\bsp	
\label{lastpage}
\end{document}